\documentstyle[aps,pra,eqsecnum,tighten]{revtex}
\input{epsf}

\def\ad{^\dagger }
\def\bt{\beta }
\def\ef#1w{(\ref{e#1})}
\def\eh{\hat e}
\def\ep{\epsilon }
\def\eph{\hat \epsilon}
\def\gm{\gamma } 
\def\half{{\textstyle {1\over2}}}
\def\hpm{\hphantom{-}}
\def\Im{\hbox{Im}}
\def\lg{{\langle }}
\def\lla{\longleftarrow }
\def\llra{\longleftrightarrow }
\def\lm{\lambda }
\def\Lm{\Lambda }
\def\lpar{\left( }
\def\lra{\longrightarrow }
\def\kp{\kappa }
\def\mt{\mapsto }
\def\ot{\otimes }
\def\quart{{\textstyle {1\over4}}}
\def\ra{{\rightarrow }}
\def\rf{\rho_{\cal F}}
\def\rpar{\right) }
\def\rg{{\rangle }}
\def\sg{\sigma }
\def\sign{\hbox{sign}}
\def\srt{\sqrt{2}}

\def\tp{^\top}
\def\Re{\hbox{Re}}
\def\Vh{\hat V}
\def\Tr{\hbox{Tr}}

\def\A{{\cal A}}

\def\B{{\cal B}}
\def\C{{\cal C}}
\def\D{{\cal D}}
\def\E{{\cal E}}
\def\F{{\cal F}}
\def\G{{\cal G}}
\def\H{{\cal H}}
\def\Q{{\cal Q}}

\def\bold#1{\mbox{\boldmath $#1$}}
\def\bb{\bold{b}}

\def\bbr{\bold{\breve b}}
\def\Bb{\bold{B}}

\def\cb{\bold{c}}
\def\cbr{\bold{\breve c}}
\def\Cb{\bold{C}}

\def\cdotb{\bold{\cdot}}
\def\delbb{\bold{\delta}^B}
\def\delbc{\bold{\delta}^C}
\def\fb{\bold{f}}
\def\gb{\bold{g}}
\def\hb{\bold{h}}
\def\Jb{\bold{J}}
\def\Kb{\bold{K}}
\def\mhb{\bold{\hat m}}

\def\Mhb{\bold{\hat M}}

\def\Pb{\bold{P}}
\def\Ppb{\bold{P'}}
\def\Pppb{\bold{P''}}
\def\rb{\bold{r}}

\def\sb{\bold{s}}
\def\sgb{\bold{\sigma }}
\def\tb{\bold{t}}
\def\xb{\bold{x}}
\def\xhb{\bold{\hat x}}

\begin{document}

\title{Optimal Copying of One Quantum Bit}

\author{Chi-Sheng Niu\thanks{Electronic mail: cn28+@andrew.cmu.edu} and
Robert B. Griffiths\thanks{Electronic mail: rgrif@cmu.edu}\\ Department of
Physics, Carnegie Mellon University,\\ Pittsburgh, PA 15213}

\maketitle

\begin{abstract}
	A quantum copying machine producing two (in general non-identical)
copies of an arbitrary input state of a two-dimensional Hilbert space (qubit)
is studied using a quality measure based on distinguishability of states,
rather than fidelity.  The problem of producing optimal copies is investigated
with the help of a Bloch sphere representation, and shown to have a
well-defined solution, including cases in which the two copies have unequal
quality, or the quality depends upon the input state (is ``anisotropic'' in
Bloch sphere language), or both.  A simple quantum circuit yields the optimal
copying machine.  With a suitable choice of parameters it becomes an optimal
eavesdropping machine for some versions of quantum cryptography, or reproduces
the Bu\v zek and Hillery result for isotropic copies.
\end{abstract}


			\section{Introduction}
\label{s1}
	
	The no-cloning theorem \cite{r1,r2} is one of the most important
features which distinguishes quantum from classical information theories.
There is no difficulty making an arbitrary number of copies of any type of
information which arrives over a classical channel.  However, to copy or
``clone'' the information which arrives over a quantum channel is not possible
without producing errors if it is encoded in terms of nonorthogonal quantum
states.  This fact is the basis for various schemes of quantum cryptography
\cite{r3}, since the attempt of an eavesdropper to tap into a quantum
channel and duplicate the information in it will result in errors detectable by
the legitimate users.

	Even in cases in which perfect copies are excluded because of the
no-cloning theorem, it is possible to produce imperfect copies which are better
than random noise.  An optimal copying machine is one in which, by means of a
suitable unitary transformation, a certain number of copies of the original
information are produced with the smallest number of errors, or minimal amount
of noise.  Optimality in this sense will obviously depend upon the ensemble of
input states, the number of copies, and the measure employed for noise or error
rate.  Obtaining quantitative estimates of what is possible for optimal copying
is a significant problem in quantum information theory, and it is to this
problem that the present paper is addressed.

	A number of previous studies have used various input ensembles and
measures of quality or of noise. Bu\v zek and Hillery discussed an
``isotropic'', or ``universal'', copying machine \cite{r4}, which produces two
identical copies from one qubit (two-state system), with copy quality
independent of the input state.  This copying machine was later proved by
{Bru\ss} {\it et al.} \cite{r5} to be optimal among all possible isotropic
machines, if the measure of quality is the fidelity between the input and the
output: that is, the probability that the output will be measured to be in the
same state as the input.  Gisin and Massar \cite{r6} considered a case in which
$M$ identical copies are generated from $N$ identical qubits $(M>N)$, and found
that the average fidelity reaches its maximum when the machine is isotropic.
Bru\ss, Ekert, and Macchiavello \cite{r7} found an interesting connection
between the optimal isotropic copying process and quantum state estimation by
measurement.  Later, the optimal $M$-to-$N$ problem was generalized by Werner
\cite{r8} to systems of arbitrary dimension.  Various quantum circuits
implementing some of the isotropic copying machines mentioned above are
presented in \cite{r9,r10}.  Bru\ss {\it et al.} \cite{r5} also considered an
input ensemble of two non-orthogonal states, and found that the optimal copying
process requires no ancillary qubits, unlike the case of isotropic copying.

	In contrast to the work just discussed, we will use a quality measure
based upon the distinguishability of the output states, rather than the
fidelity, and will consider {\it non-identical} and {\it anisotropic} copies
for which the copying quality can depend upon the input state. This is
motivated by a study of eavesdropping in quantum cryptography \cite{r11}.
Eavesdropping is, in essence, a matter of producing imperfect ``copies'' of the
input signal in some channel in a way which enables the eavesdropper to
distinguish the different states in the original information, while at the same
time perturbing it as little as possible.  Thus distinguishability is more
important to the eavesdropper than whether his copies resemble the original
states.  In addition, the amount of information obtained by the legitimate
receiver and by the eavesdropper are generally different, and can depend on the
input states.  Since the problem of optimizing a copy machine for completely
general inputs and outputs is quite difficult, we restrict ourselves to the
case of only two copies of one qubit.  Even with this restriction, the
mathematical structure is nontrivial.  Recently Cerf \cite{r12} explored this
problem by investigating a family of ``Pauli cloning machines.'' He derived a
no-cloning inequality for isotropic copies, and established a bound for the
quantum capacity of the corresponding  ``Pauli channel.''

	We introduce the background material needed for later sections in
Sec.~\ref{s2}, including a formalism for the general interaction of one qubit
with another quantum system, a set of superoperators, and the Bloch sphere
representation.  In Sec.~\ref{s3}, the {\it quality function} of an output
qubit is defined, and it is expressed using the Bloch sphere representation;
this is the measure we employ to define optimal copying.  Given the quality of
one copy, the best quality for the other is determined by a map $G$, which is
derived in Sec.~\ref{s4} for a special case.  However, this special case is as
good or better than any other copying scheme, as shown in Sec.~\ref{s5}.
Section \ref{s6} discusses a number of properties of the map $G$, the
conditions under which a pair of copies produced are actually optimal, and
particular examples that are useful in revealing relations between the copying
qualities.  In Sec.~\ref{s7} we introduce two quantum circuits implementing the
optimal copying process, and show how they can be turned into an eavesdropping
machine or a universal copying machine.  Finally, the results are summarized
and some open questions are noted in Sec.~\ref{s8}.  Various mathematical
details are placed in Appendices A--D.

		\section{General formulation of the copying problem}
\label{s2}

	We define the copying problem in the following way.  The qubit which is
to be copied is a state in a two-dimensional Hilbert space $\A$ with orthogonal
basis $|0\rg$ and $|1\rg$. The copy machine corresponds to a unitary
transformation $U$ on
\begin{equation}
  \H=\A\ot\E = \B\ot\E \ =\B\ot\C\ot\D,
\label{e2.1}
\end{equation}
where we suppose that the copies lie in the two dimensional spaces $\B$ and
$\C$ after the action of $U$, and the space $\E=\C\otimes\D$ can have an
arbitrarily large but finite (even) dimension. The identical spaces $\A$ and
$\B$ are denoted differently in order to distinguish the input space from the
output.  In particular,
\begin{equation}
  U(|j\rg\ot |\ep\rg) = \sum_{l=0}^1 |l\rg\ot |\ep^j_l\rg,
\label{e2.2}
\end{equation}
where $|\ep\rg$ is a normalized state of $\E$, and therefore the states
$|\ep^j_l\rg$ of $\E$ satisfy
\begin{equation}
  \sum_l \lg\ep^j_l|\ep^k_l\rg = \delta_{jk}.
\label{e2.3}
\end{equation}
There is no loss of generality in assuming a pure state $|\ep\rg$ rather than a
density matrix for $\E$, since a density matrix can always be thought of as
arising from a partial trace on a space of higher dimension, and the dimension
of $\E$ is arbitrary.

	With $|\ep\rg$ fixed, define the linear map $V:\A\ra\B\ot\E$ through
\begin{equation}
  V|\alpha\rg = U(|\alpha\rg\ot |\ep\rg) = 	\sum_{l=0}^3(\sg_l
|\alpha\rg)\ot |e_l\rg,
\label{e2.4}
\end{equation}
where $\sg_0$ is the identity, and $\sg_1$, $\sg_2$, and $\sg_3$ are the usual
Pauli matrices $\sg_x$, $\sg_y$, and $\sg_z$, which in (\ref{e2.4}) should be
thought of as maps from $\A$ to $\B$, assuming some basis $|0\rg,\, |1\rg$ for
the latter; here $|0\rg$ and $|1\rg$ correspond to $S_z$ equal to $+1/2$ and
$-1/2$, respectively, and $(|0\rg+|1\rg)/\srt$ to $S_x=+1/2$ in the usual spin
half notation.  Thus if (\ref{e2.4}) is written out explicitly, it takes the
form:
\begin{eqnarray}
  V|0\rg &=& |0\rg|e_0\rg + |1\rg|e_1\rg + i |1\rg|e_2\rg + |0\rg|e_3\rg,
\nonumber \\
  V|1\rg &=& |1\rg|e_0\rg + |0\rg|e_1\rg - i |0\rg|e_2\rg - |1\rg|e_3\rg,
\label{e2.5}
\end{eqnarray}
where, following the usual convention, we omit $\ot$ if this causes no
confusion.  The four $|e_l\rg$ are vectors in $\E$ which are linear
combinations of the $|\ep^j_k\rg$, and whose inner products form a $4\times 4$
positive (i.e., non-negative eigenvalues) Hermitian matrix $E$:
\begin{equation}
  E_{lm} = \lg e_l|e_m\rg = E^*_{ml}.
\label{e2.6}
\end{equation}
Note that given a positive Hermitian matrix $E$, it is always possible to find
a set of four vectors $|e_l\rg$ such that $E_{lm}$ takes the form (\ref{e2.6}),
and this set is unique up to an arbitrary unitary transformation on $\E$.

	The condition (\ref{e2.3}) implied by the unitarity of $U$ is
equivalent to the requirement that $V$ be an isometry, or that
\begin{equation}
  V\ad V=\sum_{lm} E_{lm} \sg_l\sg_m = \sg_0,
\label{e2.7}
\end{equation}
the identity on $\A$, and thus to
\begin{eqnarray}
  \sum_{l=0}^3 E_{ll} &=& 1,
\label{e2.8} \\
 \Re(E_{0q}) &=& \Im(E_{q'q''}),
\label{e2.9}
\end{eqnarray}
where $\Re$ and $\Im$ refer to real and imaginary parts, and (\ref{e2.9})
employs a notation which we will use again and again:
\begin{equation}
  (q,q',q'') = (1,2,3) \mbox{ or } (2,3,1) \mbox{ or } (3,1,2).
\label{e2.10}
\end{equation}
That is, the triple $(q,q',q'')$ is an {\it even} permutation of the integers
1, 2, and 3.  Thus, for example, $\Re[E_{02}]=\Im[E_{31}].$

	Given $V$, the corresponding {\it superoperator} $\Vh$ which maps
operators on $\A$ to operators on $\H$ takes the form
\begin{equation}
  \Vh(\kp) = V\kp V\ad = \sum_{jk} \sg_j\kp\sg_k\ot |e_j\rg\lg e_k|.
\label{e2.11}
\end{equation}
In particular, for $l=0,1,2,3,$
\begin{equation}
  \Vh(\sg_l) = \sum_{jkm} L(jk;lm)\, \sg_m\ot |e_j\rg\lg e_k| =
\sum_m \sg_m\ot\hat F_{lm}, 
\label{e2.12}
\end{equation}
where the $\hat F_{lm}$ are a collection of 16 Hermitian operators on $\E$
defined by
\begin{equation}
  \hat F_{lm} = \sum_{jk} L(jk;lm)\, |e_j\rg\lg e_k|,
\label{e2.13}
\end{equation}
and
\begin{equation}
  L(jk;lm) = \half\Tr[\sg_j\sg_l\sg_k\sg_m]
\label{e2.14}
\end{equation}
can be thought of as a $16\times 16$ matrix, with rows labeled by $jk$ and
columns by $lm$.  Its properties are discussed in App.~\ref{aa}.  In
particular, $\half L$ is its own inverse.  When $j=k$ or when $l=m$, $L(jk;lm)$
vanishes unless the other pair is also equal:
\begin{equation}
  L(jk;ll) = \delta_{jk}\Lm(j,l),\quad L(jj,lm) = \delta_{lm}\Lm (j,l),
\label{e2.15}
\end{equation}
where $\Lm(j,l)$ is a $4\times 4$ real matrix, (\ref{ea.8}), whose square is 4
times the identity.

	The superoperators $\Vh_\B,$ $\Vh_\C$, and $\Vh_\E$, which play a
central role in the following discussion, map operators on $\A$ to operators on
$\B$, $\C$, and $\E$, respectively and are defined using partial traces of
$\Vh$:
\begin{eqnarray}
  &&\Vh_\B(\kp) = \Tr_\E[\Vh(\kp)],\quad \Vh_\E(\kp) = \Tr_\B[\Vh(\kp)],
\nonumber\\
  &&\Vh_\C(\kp) = \Tr_{\B\ot\D}[\Vh(\kp)] = \Tr_\D[\Vh_\E(\kp)]
\label{e2.16}
\end{eqnarray}
The isometry condition (\ref{e2.7}) implies that $\Vh$ preserves the trace,
\begin{equation}
  \Tr_\H[\Vh(\kp)] = \Tr_\A[\kp],
\label{e2.17}
\end{equation}
and this property is inherited by each of the superoperators $\Vh_\B,$
$\Vh_\C$, and $\Vh_\E$, with the trace over $\H$ in (\ref{e2.17}) replaced by a
trace over $\B$, $\C$, and $\E$, respectively.  (Henceforth we shall not always
indicate explicitly the space over which a trace is to be taken if it is
obvious from the context.) In addition, it is obvious from (\ref{e2.11}) that
$\Vh$ maps Hermitian operators to Hermitian operators, and this property is
also inherited by $\Vh_\B,$ $\Vh_\C$, and $\Vh_\E$.

	It is convenient to write
\begin{equation}
  \Vh_\B(\sg_l) = \sum_m B_{lm}\sg_m,
\label{e2.18}
\end{equation}
where
\begin{equation}
  B_{lm} = \Tr_\E[\hat F_{lm}] = \sum_{jk} L(lm;jk)\, E_{jk}
\label{e2.19}
\end{equation}
is a real $4\times 4$ matrix of the form:
\begin{equation}
  B=\left(\matrix{ 1 & \delta^B_1 & \delta^B_2 & \delta^B_3 \cr 0 & & & \cr 0 &
&\Bb& \cr 		 0 & & & }\right),
\label{e2.20}
\end{equation}
where columns and rows are in the order $0,1,2,3$, and $\Bb$ is a real
$3\times3$ matrix.  The form of the first (left-most) column of $B$ is
equivalent to the isometry condition (\ref{e2.7}), that is, to (\ref{e2.8})
plus (\ref{e2.9}), as can be shown by inserting in (\ref{e2.19}) the matrix
elements for the $L$ given in App.~\ref{aa}.  It implies that the superoperator
in (\ref{e2.18}) preserves the trace.  The inverse of (\ref{e2.19}), see
(\ref{ea.10}), is
\begin{equation}
  E_{jk}={\textstyle {1\over4}}\sum_{lm} L(jk;lm)\, B_{lm}.
\label{e2.21}
\end{equation}

	The vector $\delbb = (\delta^B_1,\delta^B_2,\delta^B_3)$ and the matrix
$\Bb$ in (\ref{e2.20}) can be given a geometrical interpretation using the
well-known Bloch sphere representation
\begin{equation}
  \rho=\half(I+\bold{r\cdot\sg}) = \half(\sg_0 +\sum_{j=1}^3 r_j\sg_j)
\label{e2.22}
\end{equation}
of a density matrix $\rho$ in a two-dimensional Hilbert space, where
$\rb=(r_1,r_2,r_3)$ is any real vector of length less than 1 for a mixed state,
or equal to 1 for a pure state, and $\sgb=(\sg_1,\sg_2,\sg_3)$.  Then
(\ref{e2.18}) and (\ref{e2.20}) tell us that
\begin{equation}
  \Vh_\B(\rho)=\rho'=\half(\sg_0+\bold{s\cdot\sg}),
\label{e2.23}
\end{equation}
where \cite{r13}
\begin{equation}
  \sb=\delbb+\bold{r\cdot B} = \delbb+\bold{ B\tp\cdot r},
\label{e2.24}
\end{equation}
and $\bold{B\tp}$ is the transpose of $\Bb$.  In particular, if (\ref{e2.24})
is applied to the collection of pure states, $|\rb|=1$, which form the unit
Bloch sphere of $\A$, the resulting collection of $\sb$ vectors forms an
ellipsoid in the Bloch sphere of $\B$, which we shall call the {\it $\B$
ellipsoid}, or simply {\it ellipsoid}, with center at $\delbb$.

	The polar decomposition \cite{r14} of $\Bb$ takes the form
\begin{equation}
  \Bb=\bold{P\cdot B^s},\label{e2.25}
\end{equation}
where $\Pb$ is a proper rotation, a real orthogonal $3\times 3$ matrix with
determinant 1, and $\bold{B^s}$ is a real symmetric matrix.  By diagonalizing
$\bold{B^s}$, one can always write $\Bb$ in the form
\begin{equation}
  \Bb=\bold{P'\cdot B^d\cdot P''},
\label{e2.26}
\end{equation}
where $\Ppb$ and $\Pppb$ are proper rotations, and $\bold{B^d}$ is a diagonal
matrix with (real) eigenvalues $b_1,b_2,b_3$.  The rotations $\Ppb$ and $\Pppb$
correspond to unitary transformations on $\A$ and $\B$, respectively.
Consequently, by adopting suitable orthonormal bases for $\A$ and $\B$, we can,
assuming $U$ to be given, always arrange to have the matrix $B$ in the form
\begin{equation}
  B=\left(\matrix{ 1 & \delta^B_1 & \delta^B_2 & \delta^B_3 \cr 0 & b_1 & 0 & 0
\cr 		 0 & 0 & b_2 & 0 \cr 		 0 & 0 & 0 & b_3 }\right),
\label{e2.27}
\end{equation}
parameterized by $\delbb=(\delta^B_1,\delta^B_2,\delta^B_3)$ and
$\bb=(b_1,b_2,b_3)$.  (Note that in the process of diagonalizing the general
form (\ref{e2.20}) to obtain the diagonal form (\ref{e2.27}), the values of the
$\delta^B_j$ will, in general, change.)

	Two vectors corresponding to an orthonormal basis or {\it mode} for
$\A$ are represented by opposite poles $\mhb=(m_1,m_2,m_3)$ and $-\mhb$ of the
Bloch sphere of $\A$, where we use a hat to indicate a unit vector. Under
(\ref{e2.24}) these map to opposite points of the $\B$ ellipsoid; that is, the
line connecting them passes through the center $\delbb$ of the ellipsoid. The
three principal axes of the $\B$ ellipsoid correspond to three {\it principal
modes} of $\A$, represented in the Bloch sphere of $\A$ by mutually
perpendicular unit vectors (and their negations) $\Mhb_1$, $\Mhb_2$, and
$\Mhb_3$.  If bases are chosen such that $B$ has the form (\ref{e2.27}), the
mode vectors $\Mhb_j$ are along the Cartesian axes, and correspond, in the
language of spin half, to a value of $+1/2$ for $S_x$, $S_y$, and $S_z$,
respectively.  (It is, of course, important to distinguish ``orthogonal'' with
reference to the complex Hilbert spaces $\A$ or $\B$, from ``perpendicular'' as
it refers to real three-dimensional vectors in the Bloch sphere
representation!)  Also, $|b_j|$ is the length of the $j'$th principal semi-axis
of the ellipsoid.

	Because $\Ppb$ and $\Pppb$ are proper rotations (each has determinant
1), it is not always possible to make all the $b_j$, $j=1,2,3$, in
(\ref{e2.27}) positive (non-negative).  If an even number of the $b_j$ are
negative, then $\Ppb$ and $\Pppb$ can be chosen so they are all positive, but
if an odd number are negative, the best one can do is to have two positive and
one negative.  Which of these situations occurs depends upon $U$, and turns out
to be of some significance for optimal copying.

	Since $\C$ is also a two-dimensional space, we can use a Bloch sphere
representation for $\Vh_\C$ and write the counterpart of (\ref{e2.23}) as
\begin{equation}
    \Vh_\C(\rho)=\rho''=\half(\sg_0+\bold{t\cdot\sg}),
\label{e2.28}
\end{equation}
where
\begin{equation}
  \tb=\delbc+\bold{r\cdot C} = \delbc+\bold{ C\tp\cdot r}.
\label{e2.29}
\end{equation}
The counterpart of (\ref{e2.20}) is a $4\times4$ matrix
\begin{equation}
  C=\left(\matrix{ 1 & \delta^C_1 & \delta^C_2 & \delta^C_3 \cr 0 & & & \cr 0 &
&\Cb& \cr 		 0 & & & }\right),
\label{e2.30}
\end{equation}
where $\Cb$ is a $3\times3$ matrix which can be diagonalized by the same
strategy employed earlier in the case of $\Bb$, though the principal modes of
$\A$ singled out by this construction will, in general, not be the same as
those corresponding to the $\B$ ellipsoid, so we label the perpendicular unit
vectors with a prime: $\Mhb'_1$, $\Mhb'_2$, and $\Mhb'_3$.  We shall denote by
$\cb=(c_1,c_2,c_3)$ the values on the diagonal when $\Cb$ is diagonalized;
hence $|c_j|$ is the length of the $j'$th principal semi-axis of the $\C$
ellipsoid.

	The possible values of $\bb$ and $\delbb$ are constrained by the fact
that, whatever $U$ may be, the matrix $E$, which determines $B$ through
(\ref{e2.19}), and is determined by $B$ through the inverse relationship
(\ref{e2.21}), satisfies the isometry condition (\ref{e2.8}) and (\ref{e2.9})
and has non-negative eigenvalues.  One consequence is the fact that the $\B$
ellipsoid must always lie inside the unit Bloch sphere of $\B$.  These
constraints are found explicity for the case $\delbb=\bold{0}$ in Sec.~\ref{s4}
below, but we do not know their form in general.  There are, of course,
identical constraints on the possible values of $\cb$ and $\delbc$

		\section{Error rates and quality function}
\label{s3}

	Consider the problem of a general channel of the type introduced in
Sec.~\ref{s2}, from a qubit $\A$ to a Hilbert space $\F$ (which could be $\B$
or $\C$ or $\E$), described by a superoperator $\Vh_\F$.  Suppose that symbols
$a_1$ and $a_2$ are encoded in basis states $\mhb$ and $-\mhb$ in the Bloch
sphere representation of $\A$, and the corresponding states are mapped by
$\Vh_\F$ to the two density matrices $\rf^1$ and $\rf^2$ in $\F$.  Now suppose
that every time a signal $a_1$ or $a_2$ is sent, a measurement is carried out
on $\F$ corresponding to a decomposition of the identity
\begin{equation}
  I_\F =\sum_k F_k,
\label{e3.1}
\end{equation}
where the $\{F_k\}$ are projectors, or, more generally, positive operators
\cite{r15}. Further suppose that if the result of this measurement is $f_k$
corresponding to $F_k$, an estimate of the original symbol is made according to
a fixed guessing function%
\footnote{One can employ values for $g(a_j|f_k)$ between 0 and 1, provided
(\ref{e3.2}) is satisfied, for a random guessing process in which $g(a_j|f_k)$
is the probability of guessing $a_j$ given $f_k$.  However, one can show that
this more general procedure does not result in a smaller error rate, so we
shall not consider it further.}
$g(a_j|f_k)$ which is either 0 or 1, and satisfies
\begin{equation}
  \sum_j g(a_j|f_k)=1.
\label{e3.2}
\end{equation}
That is, if $g(a_2|f_1)=1$, then if $f_1$ is measured, one guesses that $a_2$
was sent.  If $a_j$ is sent with probability $p_j$, one can show that the
average error rate (guessing $a_1$ when $a_2$ was sent, or vice versa) using
this procedure is bounded below, see \cite{r16}, by the (achievable) minimum
error rate
\begin{equation}
  \ep_{min}=\half\left[ 1-\Tr\Bigl(|p_1\rf^1 -p_2\rf^2|\Bigr)\right],
\label{e3.3}
\end{equation}
where the absolute value $|C|$ of a Hermitian operator $C$ is the operator
obtained by replacing the eigenvalues of $C$ with their absolute values in its
spectral decomposition. Motivated by (\ref{e3.3}), we define the {\it quality
factor} $\Q_\F(\mhb)$ or ``distinguishability'' for mode $\mhb$ as
\begin{equation}
  \Q_\F(\mhb)=\half\Tr\lpar |\rf^1-\rf^2|\rpar = 	\Tr\Bigl(
|\Omega_\F(\mhb)| \Bigr),
\label{e3.4}
\end{equation}
where
\begin{equation}
 \Omega_\F(\mhb)= \Vh_\F(\half\bold{\hat m\cdot\sg}) = 	\half\bold{\hat
m\cdot}\Vh_\F(\sgb).
\label{e3.5}
\end{equation}
Thus $\Q_\F(\mhb)$ is a number between 0 and 1, equal to $1-2\ep_{min}$ in the
case in which $p_1=p_2=1/2$. A quality factor of 1 means the two signals can be
perfectly distinguished, while 0 means that measurements are no more effective
than random guesses.

	The quality factor as a function of mode, $\Q_\F(\mhb)$, will be called
the {\it quality function}.  Knowledge of this function provides a general
measure of ``distinguishability'' for pairs of non-orthogonal as well as
orthogonal signals in $\A$, for if $a_1$ and $a_2$ correspond to density
matrices
\begin{equation}
  a_j \leftrightarrow \rho_{\cal A}^j = \half\lpar I+\xb^j\bold{\cdot\sg}\rpar
\label{e3.6}
\end{equation}
in $\A$, the distinguishability of the corresponding signals in $\F$, compare
(\ref{e3.4}), is
\begin{equation}
  \half\Tr\Bigl( |\Vh_\F(\rho_{\cal A}^1)-\Vh_\F(\rho_{\cal A}^2)| \Bigr)
={\textstyle {1\over4}}\, x\,\Tr\Bigl( |\bold{x\cdot} \Vh_\F(\sgb)|\Bigr)
=\half\, x\, \Q_\F(\xhb),
\label{e3.7}
\end{equation}
where $\xhb$ is a unit vector in the direction of $\xb^1-\xb^2$, and
$x=|\xb^1-\xb^2|$.

	The quality factor $\Q_\F(\mhb)$ for mode $\mhb$ is, (\ref{e3.4}), the
sum of the absolute values of the eigenvalues of $\Omega_\F(\mhb)$.  When
$\F=\H$, $\Vh_\F=\Vh$, and since $\Vh$ is an isometry, $\Vh(\half\bold{\hat
m\cdot\sg})$ has two non-zero eigenvalues, $+1/2$ and $-1/2$, the same as those
of $\half\bold{\hat m\cdot\sg}$.  Hence $\Q_\H(\mhb)=1$, independent of $\mhb$.
When $\F=\B$ or $\E$, it is helpful to note that $\Omega_\B$ and $\Omega_\E$
can be obtained from partial traces of
\begin{equation}
  \Omega_\H(\mhb) = \half\sum_{q=1}^3 m_q \sum_{l=0}^3 \sg_l\ot\hat F_{ql},
\label{e3.8}
\end{equation}
see (\ref{e2.12}).  In particular, one has
\begin{equation}
  \Omega_\B(\mhb)=\Tr_\E[\Omega_\H(\mhb)]= \half\bold{\hat m\cdot B\cdot\sg},
\label{e3.9}
\end{equation} 
since $B_{lm}$, (\ref{e2.19}), is the trace of $\hat F_{lm}$, and $B_{q0}=0$,
see (\ref{e2.20}).  The eigenvalues of $\Omega_\B(\mhb)$ are
$\pm\half|\bold{\hat m \cdot B}|$, where $\bold{\hat m
\cdot  B}$ (see the discussion in Sec.~\ref{s2}) is the vector from the
center of the ellipsoid to the point on its surface which is the image, under
$\Vh_\B$, of $\mhb$, and $|\bold{\hat m \cdot B}|$ is its length.  The same
argument applies when $\F$ is $\C$, and thus we have
\begin{equation}
  \Q_\B(\mhb) = |\bold{\hat m \cdot B}|,\quad \Q_\C(\mhb) = |\bold{\hat m \cdot
C}|.
\label{e3.10}
\end{equation}
If we write $\mhb$ in terms of the principal modes of $\A$ relative to the $\B$
and $\C$ ellipsoids,
\begin{equation}
  \mhb=\sum_{j=1}^3 \mu_j\Mhb_j=\sum_{j=1}^3 \mu'_j\Mhb'_j,
\label{e3.11}
\end{equation}
respectively, the two quality functions can be written explicitly as:
\begin{equation}
  \Q_\B(\mhb)=\surd\bigl\{\sum_{j=1}^3 (b_j \mu_j)^2\bigr\},\quad
\Q_\C(\mhb)=\surd\bigl\{\sum_{j=1}^3 (c_j \mu'_j)^2\bigr\},
\label{e3.12}
\end{equation}
where the $b_j$ and the $c_j$ are the diagonal elements of the diagonalized
matrices $\Bb$ and $\Cb$, and represent in each case the length of the $j'$th
principal semi-axis of the ellipsoid.

	In the case $\F=\E$, we have
\begin{equation}
  \Omega_\E(\mhb)= \sum_{q=1}^3 m_q \hat F_{q0}
\label{e3.13}
\end{equation}
upon tracing (\ref{e3.8}) over $\B$.  There is no simple expression (known to
us) for the spectrum of $\Omega_\E(\mhb)$, and we cannot represent the quality
function $\Q_\E(\mhb)$ using a Bloch sphere.  Nonetheless, it is worth noting
that $\Q_\E(\mhb)$ is completely determined if the superoperator $\Vh_\B$, or,
equivalently, the matrix $B$, is given.  That is because $B$ determines $E$,
(\ref{e2.21}), and as noted below Eq.~(\ref{e2.6}), it thus determines $\hat
F_{lm}$, see (\ref{e2.13}), up to a unitary transformation on $\E$.  As a
consequence, this transformation cannot change $\Q_\E(\mhb)$ since it leaves
the spectrum of $\Omega_\E(\mhb)$ the same.  On the other hand, $\Q_\C(\mhb)$
is not uniquely determined by $\Vh_\B$, because there are many ways in which
$\E$ can be expressed as a tensor product $\C\ot\D$.  Nonetheless, it is clear
that
\begin{equation}
   \Q_\C(\mhb) \leq \Q_\E(\mhb),
\label{e3.14}
\end{equation}
because measurements on $\C$ alone clearly cannot do better at distinguishing
signals than those carried out on the full space $\E$. A formal proof of
(\ref{e3.14}) is obtained by noting that
\begin{equation}
   \Omega_\C(\mhb)=\Tr_\D[\Omega_\E(\mhb)],
\label{e3.15}
\end{equation}
and using theorem 3 of App.~B.

		\section{Optimal Copying: Centered Ellipsoid}
\label{s4}

	We approach the problem of optimal copying in the following way.
Suppose a copying machine produces one copy in the $\B$ output channel
characterized by a quality function $\Q_\B(\mhb)$ which, as noted in
Sec.~\ref{s3}, is completely specified by a set of principal modes in $\A$
together with the semi-axes of the corresponding ellipsoid in the $\B$ Bloch
sphere, the absolute values of $b_1$, $b_2$, and $b_3$.  We then ask, given
$\bb=(b_1,b_2,b_3)$, what is the best possible copy which can be produced in a
separate one qubit (two-dimensional) output channel $\C$; that is, what is the
best or largest quality function $\Q_\C(\mhb)$?

	It is not obvious at the outset that this question has a well-defined
answer, for one can certainly imagine that with a fixed $\bb$, one type of copy
machine might produce an optimal quality in $\C$ for mode $\mhb$, and a
different machine would be needed to make the best copies for a different mode
$\mhb'$.  But we will show that for a given $\bb$ there is, indeed, a single
machine which produces the best possible copies in $\C$, as measured by
$\Q_\C$, for every $\mhb$.  That is, given $\bb$, there is a largest ellipsoid
in the Bloch sphere of $\C$, characterized by semi-axes $c_1$, $c_2$, and
$c_3$, with
\begin{equation}
  \cb = (c_1,c_2,c_3) = G(\bb),
\label{e4.1}
\end{equation}
a well-defined function of $\bb$.  The construction, incidentally, yields
non-negative $c_j$.

	In the present section, the optimization map $G$ will be derived by
considering the special case of an ellipsoid centered in the Bloch sphere of
$\B$, that is, $\delbb=0$ in (\ref{e2.27}).  That considering this special case
is actually sufficient will be shown in Sec.~\ref{s5}, where we use a concavity
property of the quality function to show that the quality cannot be further
improved by using an ellipsoid which is {\it not} centered in the Bloch sphere
of $\B$.  However, given the pair $\bb$ and $\cb=G(\bb)$, it is not at all
obvious that the corresponding copy machine is optimal, since there might be a
machine which produces better copies both in the $\B$ channel and in the $\C$
channel.  This possibility is studied below in Sec.~\ref{s6}, where we also
work out certain properties of $G$.

	If the $|e_l\rg$ in (\ref{e2.4}) are mutually orthogonal, the matrix
$E_{lm}$ of inner products, (\ref{e2.6}) is diagonal, and can be written in the
form
\begin{equation}
  E_{lm} = \beta^2_l\delta_{lm},
\label{e4.2}
\end{equation}
where the $\beta_l$ are non-negative numbers. Further, we can choose an
orthonormal basis $\{|\hat e_l\rg\}$ in $\E$ such that
\begin{equation}
  |e_l\rg = \beta_l |\hat e_l\rg,
\label{e4.3}
\end{equation}
The symbol $\beta$ without a subscript will be used to denote the four vector
$(\beta_0,\beta_1,\beta_2,\beta_3)$.  Because of (\ref{e2.15}), if $E$ is
diagonal, the matrix $B$, (\ref{e2.19}), is also diagonal,
\begin{equation}
  B_{lm} = b_l\delta_{lm},
\label{e4.4}
\end{equation}
and we will use $b$ without a subscript for the four vector of diagonal
elements:
\begin{equation}
  b= (b_0,b_1,b_2,b_3) = (1,\bb).
\label{e4.5}
\end{equation}
Using (\ref{e2.19}), (\ref{e2.15}), and (\ref{e4.2}), we can express the
elements of $b$ in terms of the $\bt_j$ through
\begin{equation}
  b=\Lm\cdot\beta^2,\quad \beta^2 = \quart\Lm\cdot b,
\label{e4.6}
\end{equation}
where the dot denotes matrix multiplication, and
\begin{equation}
  \bt^2=(\bt_0^2,\bt_1^2,\bt_2^2,\bt_3^2)
\label{e4.7}
\end{equation}
is a four vector whose components are the squares of those of $\bt$.  If
(\ref{e4.6}) is written out explicitly one finds:
\begin{eqnarray}
  &&b_0 = 1 = \sum_{j=0}^3 \beta_j^2
\label{e4.8}\\
  &&b_q = \beta_0^2 + \beta_q^2 - \beta_{q'}^2 - \beta_{q''}^2 	\hbox{ for }
q=1,2,3.
\label{e4.9}
\end{eqnarray}

	Since the off-diagonal $\delbb$ in (\ref{e2.20}) is zero, (\ref{e4.4}),
the ellipsoid is centered, and the fact that $\Bb$ is diagonal means that the
principal axes of the ellipsoid coincide with the Cartesian axes of the Bloch
sphere.  Note that the property last mentioned can always be achieved by
choosing an appropriate orthonormal basis in $\B$, so that there is no loss of
generality in assuming, if the ellipsoid is centered, that the $B$ matrix is
diagonal, of the form (\ref{e4.4}).  But in that case (\ref{e2.21})---note
(\ref{e2.15})---tells us that the matrix $E$ is diagonal, so that the $|e_l\rg$
must be mutually orthogonal.  Thus by considering all mutually orthogonal sets
of $|e_l\rg$, we take care of all cases in which the $\B$ ellipsoid is
centered.
	The fact that each component of $\bt^2$ in (\ref{e4.6}) is
non-negative, together with $b_0=1$, places a set of four constraints on the
components of $\bb$:
\begin{equation}
  \sum_{p=1}^3 b_p \geq -1;\quad b_q+b_{q'} \leq 1 + b_{q''}
\hbox{ for } q=1,2,3.
\label{e4.10}
\end{equation}
These are the necessary and sufficient conditions for $\bb$ to be physically
possible when $\delbb=0$, and they specify that it lies within a tetrahedron
with vertices
\begin{equation}
  (1,-1,-1),\ (-1,1,-1),\ (-1,-1,1),\ (1,1,1).
\label{e4.11}
\end{equation}

As noted at the end of Sec.~\ref{s3}, $\bb$ in (\ref{e4.9}) determines
$\Q_\E(\mhb)$ but not $\Q_\C(\mhb)$, which depends on the tensor product
structure $\C\otimes\D$ of $\E$.  One choice that maximizes $\Q_\C(\mhb)$ is
the set of basis vectors $|\eh_l\rg$ with
\begin{eqnarray}
  \srt|\eh_0\rg = |00\rg +|11\rg, &\quad& \srt|\eh_1\rg = |01\rg +|10\rg,
\nonumber\\
  \srt|\eh_2\rg = -i|01\rg +i|10\rg, &\quad& \srt|\eh_3\rg = |00\rg -|11\rg,
\label{e4.12}
\end{eqnarray}
where orthonormal basis vectors for the two-dimensional space $\C$ are denoted
by $|0\rg$ and $|1\rg$, the same notation is used for the basis of $\D$, and in
the kets on the right sides in (\ref{e4.12}) the $\C$ label is to the left of
the $\D$ label, thus $|cd\rg$.  The space $\D$ is two-dimensional, but could be
higher-dimensional with additional basis vectors $|2\rg$, $|3\rg$, ...etc.,
which do not appear in (\ref{e4.12}).  Substituting (\ref{e4.12}) into
(\ref{e4.3}), and (\ref{e4.3}) into (\ref{e2.5}), we obtain the explicit
expressions
\begin{eqnarray}
  \srt V|0\rg &=& (\bt_0+\bt_3)|000\rg + (\bt_0-\bt_3)|011\rg
+(\bt_1+\bt_2)|101\rg + (\bt_1-\bt_2)|110\rg,
\nonumber\\
   \srt V|1\rg &=& (\bt_0-\bt_3)|100\rg + (\bt_0+\bt_3)|111\rg
+(\bt_1-\bt_2)|001\rg + (\bt_1+\bt_2)|010\rg,
\label{e4.13}
\end{eqnarray}
where the labels in the kets on the right side are in the order $\B$, $\C$,
$\D$, that is to say, $|bcd\rg$.  Note that in Bloch sphere notation,
(\ref{e2.22}), $|0\rg$ corresponds to $\rb=(0,0,1)$, $|1\rg$ to $\rb=(0,0,-1)$,
$(|0\rg+|1\rg)/\srt$ to $\rb=(1,0,0)$, etc.  Keeping this correspondence in
mind, it is easy to check that the superoperator $\Vh_\B$ based upon
(\ref{e4.13}) maps $\rb=(1,0,0)$, $(0,1,0)$, and $(0,0,1)$ to $\sb=(b_1,0,0)$,
$(0,b_2,0)$, and $(0,0,b_3)$, respectively, where the $b_q$ are given by
(\ref{e4.9}).

	We now define a four vector $\gm=(\gm_0,\gm_1,\gm_2,\gm_3)$ in terms of
$\bt$ through:
\begin{equation}
  \gm = \half \Lm\cdot\bt,\quad \bt=\half\Lm\cdot\gm,
\label{e4.14}
\end{equation}
and rewrite (\ref{e4.13}) in terms of the $\gm_j$ as:
\begin{eqnarray}
  \srt V|0\rg &=& (\gm_0+\gm_3)|000\rg + (\gm_1+\gm_2)|011\rg
+(\gm_0-\gm_3)|101\rg + (\gm_1-\gm_2)|110\rg,
\nonumber\\
   \srt V|1\rg &=& (\gm_1+\gm_2)|100\rg + (\gm_0+\gm_3)|111\rg
+(\gm_1-\gm_2)|001\rg + (\gm_0-\gm_3)|010\rg.
\label{e4.15}
\end{eqnarray}
Note that if the first two labels, the $\B$ and $\C$ bits, are interchanged on
the right side of (\ref{e4.13}) and, at the same time, each $\bt_j$ is replaced
by $\gm_j$, the result is (\ref{e4.15}); that is to say, {\it the $\gm_j$ play
precisely the same role for the space $\C$ as the $\bt_j$ for the space $\B$.}
Consequently, we can immediately conclude that the ellipsoid in the Bloch
sphere of the $\C$ channel, produced by the action of $\Vh_\C$, is centered and
has principal semi-axes $c_1$, $c_2$, and $c_3$ lying parallel to the three
Cartesian axes, with
\begin{equation}
  c=(c_0,c_1,c_2,c_3) = (1,\cb) = \Lm\cdot\gm^2,\quad \gm^2 = \quart\Lm\cdot c,
\label{e4.16}
\end{equation}
the counterpart of (\ref{e4.6}), and $\gm^2$ the four vector whose $j'$th
component is $\gm_j^2$.  Inserting (\ref{e4.14}) in (\ref{e4.16}), one obtains
explicit expressions for the $c_q$ with $q > 0$:
\begin{equation}
    c_q = 2(\bt_0\bt_q+\bt_{q'}\bt_{q''}),
\label{e4.17}
\end{equation}
where $q'$ and $q''$ follow the notation of (\ref{e2.10}).  In addition, since
the principal axes of the ellipsoids in the $\B$ and $\C$ Bloch spheres
correspond to the same principal modes in $\A$, we can omit the primes in
(\ref{e3.11}) and (\ref{e3.12}). Thus, in particular, the quality function for
$\C$ is given by
\begin{equation}
  \Q_\C(\mhb)=\surd\bigl\{\sum_{j=1}^3 (c_j \mu_j)^2\bigr\}.
\label{e4.18}
\end{equation}

	The next step is to show that (\ref{e4.18}) is, indeed, the {\it
optimal} quality function, given $b_1$, $b_2$, and $b_3$.  To demonstrate that
this is the case, we shall show that (\ref{e3.14}) is satisfied as an equality:
$\Q_\C(\mhb)=\Q_\E(\mhb)$. In order to evaluate $\Q_\E(\mhb)$, one needs to
find the eigenvalues of the Hermitian operator $\Omega_\E(\mhb)=
\Vh_\E(\half\bold{\hat m\cdot\sg})$, whose matrix
\begin{equation}
  \lg\eh_l |\Omega_\E(\mhb) | \eh_m\rg = \sum_{j=1}^3 \mu_j L(lm;j0)\bt_l\bt_m,
\label{e4.19}
\end{equation}
when written out explicitly using (\ref{e3.13}) and (\ref{e2.13}), takes the
form:
\begin{equation}
\Omega_\E=\left(\matrix{ 
	0 & \mu_1\bt_0\bt_1 & \mu_2\bt_0\bt_2 & \mu_3\bt_0\bt_3 \cr
\mu_1\bt_0\bt_1 & 0 & i\mu_3\bt_1\bt_2 & -i\mu_2\bt_1\bt_3 \cr \mu_2\bt_0\bt_2
& -i\mu_3\bt_1\bt_2 & 0 & i\mu_1\bt_2\bt_3 \cr \mu_3\bt_0\bt_3 &
i\mu_2\bt_1\bt_3 & -i\mu_1\bt_2\bt_3 & 0 		}\right).
\label{e4.20}
\end{equation}
In (\ref{e4.19}) and (\ref{e4.20}) we have replaced each component $m_q$ of
$\mhb$ by the corresponding $\mu_q$, since---see (\ref{e3.11})---the principal
modes $\Mhb_j$ are along the coordinate axes.

	The characteristic polynomial of (\ref{e4.20}) is
\begin{equation}
  \det(\Omega_\E-\lm I) = \lm^4 - \Gamma \lm^2 + \Delta^2,
\label{e4.21}
\end{equation}
where---once again employing the notation in (\ref{e2.10})---
\begin{equation}
  \Gamma = \sum_{q=1}^3 \mu_q^2(\bt_0^2\bt_q^2 + \bt_{q'}^2\bt_{q''}^2), \quad
\Delta = \bt_0\bt_1\bt_2\bt_3(\sum_{q=1}^3 \mu_q^2).
\label{e4.22}
\end{equation}
Evidently, the eigenvalues of $\Q_\E$ occur in pairs $\pm\lm_1$, $\pm\lm_2$,
where we assume that $\lm_1$ and $\lm_2$ are positive.  Thus $\Tr(|\Omega_\E|)$
is $2(\lm_1+\lm_2)$, and one finds, after a bit of algebra, that
\begin{equation}
  \Q_\E(\mhb)=\Tr\Bigl( |\Omega_\E(\mhb)| \Bigr) 	=\sqrt{\Gamma +
2|\Delta|} 	=\surd\bigl\{\sum_{j=1}^3 \left(\breve c_j
\mu_j\right)^2\bigr\}
\label{e4.23}
\end{equation}
where
\begin{equation}
  \breve c_q = 2(|\bt_0\bt_q|+|\bt_{q'}\bt_{q''}|).
\label{e4.24}
\end{equation}
Absolute value signs have been inserted because even though we have assumed
that $\bt_j$ are non-negative in (\ref{e4.2}), we could use negative values in
(\ref{e4.19}) or (\ref{e4.20}), and it is clear from (\ref{e4.21}) and
(\ref{e4.22}) that the spectrum of $\Omega_\E$ depends only on the squares of
the $\bt_j$.  On the other hand, as long as the $\bt_j$ are non-negative, one
sees, by comparing (\ref{e4.17}) with (\ref{e4.24}), and (\ref{e4.18}) with
(\ref{e4.23}), that $\Q_\C(\mhb)$ and $\Q_\E(\mhb)$ are, indeed, identical.
Hence, if the ellipsoid is centered in the $\B$ Bloch sphere, and the values
$b_1$, $b_2$, and $b_3$ are specified, the construction associated with
(\ref{e4.12}) or, equivalently, (\ref{e4.13}), with the $\bt_j$ assumed to be
non-negative, yields copies in the $\C$ channel of the highest possible
quality, whatever mode is chosen; see the discussion associated with
(\ref{e3.14}).

		\section{Non-Centered Case: Symmetry and Concavity}
\label{s5}

	\subsection{Centered Case is Optimal}
\label{s5a}

	The problem of optimal copying has been solved in Sec.~\ref{s4} in the
sense that if the ellipsoid is {\it centered} in the $\B$ channel and has
principal semi-axes $\bb=(b_1,b_2,b_3)$, then the optimal quality for copies in
the $\C$ channel can be achieved using a centered ellipsoid with principal
semi-axes $\cb=(c_1,c_2,c_3)=G(\bb)$.  Could the copies in $\C$ be be further
improved by employing a copy machine for which the $\B$ ellipsoid is {\it not}
centered in the Bloch sphere?  That it cannot follows from the argument given
below, which is based upon some properties of of $\Q_\E(\mhb)$ which are
themselves not without interest, and are derived in Sec.~\ref{s5b} and
Sec.~\ref{s5c} below.

	As noted at the end of Sec.~\ref{s3}, $\Q_\E(\mhb)$ for a fixed mode
$\mhb$ is completely determined by the $4\times4$ matrix $B$ of (\ref{e2.20}).
Since for the following argument we can assume that $\mhb$ is held fixed, it
will be convenient to denote the dependence of $\Q_\E(\mhb)$ on $B$ by
$\Q_\E[B]$ or, when B has the diagonal form (\ref{e2.27}), by
$\Q_\E[\bb,\delbb]$, suppressing the reference to $\mhb$.  Here $\bb$ is the
set of ellipsoid semi-axes and $\delbb$ is the displacement of the ellipsoid
from the center of the Bloch sphere.

	The first property of $\Q_\E[B]$ which is of interest is the symmetry
\begin{equation}
  \Q_\E[\bb,\delbb]= \Q_\E[\bb,-\delbb]
\label{e5.1}
\end{equation}
proved in Sec.~\ref{s5b}.  The second, proved in Sec.~\ref{s5c}, is that
$\Q_\E[B]$ is a concave function in the sense that, whenever $B^1$ and $B^2$
are acceptable $B$ matrices, that is, generated by isometries of the form
(\ref{e2.4}), and $p_1$ and $p_2$ are positive numbers summing to 1,
$p_1B^1+p_2B^2$ is also an acceptable matrix, and
\begin{equation}
  \Q_\E[p_1B^1+p_2B^2] \geq p_1\Q_\E[B^1] + p_2\Q_\E[B^2].
\label{e5.2}
\end{equation}

	Given the symmetry and concavity properties just mentioned, the fact
that an ``off center'' copying machine cannot improve upon the ``centered''
version of Sec.~\ref{s4} can be shown in the follow way.  Let the copy machine
produce an ellipsoid $[\bb,\delbb]$ in the $\B$ channel, and consider a
particular mode $\mhb$ for which this machine produces a copy quality
$\Q_\C(\mhb)$ in the $\C$ channel; of course, $\Q_\C(\mhb)$ cannot,
(\ref{e3.14}), exceed $\Q_\E(\mhb)$, which we denote by $\Q_\E[\bb,\delbb]$.
Then, with an obvious choice for $B^1$ and $B^2$, and with $p_1=p_2=1/2$,
(\ref{e5.2}) tells us that
\begin{equation}
  \Q_\E[\bb,0] \geq (\Q_\E[\bb,\delbb] + \Q_\E[\bb,-\delbb])/2 	=
\Q_\E[\bb,\delbb],
\label{e5.3}
\end{equation}
where we have made use of the symmetry (\ref{e5.1}). In Sec.~\ref{s4} we showed
that for a centered ellipsoid $[\bb,0]$ it is possible to construct an optimal
copying machine which in the $\C$ channel produces copies of quality
\begin{equation}
  \Q_\C^{opt}(\mhb) = \Q_\E[\bb,0],
\label{e5.4}
\end{equation}
and, consequently, the $\C$ quality for the ``off center'' machine is bounded
by
\begin{equation}
  \Q_\C(\mhb) \leq \Q_\C^{opt}(\mhb).
\label{e5.5}
\end{equation}
Since this is true for any mode $\mhb$, no improvement over the ``centered''
machine of Sec.~\ref{s4} is possible.

	\subsection{Proof of the Symmetry Property}
\label{s5b}

	The result (\ref{e5.1}) can be established as follows.  Let $V$ in
(\ref{e2.4}) be the isometry which gives rise to an ellipsoid $[\bb,\delbb]$,
and define a second isometry
\begin{equation}
  V'|\alpha\rg = U(|\alpha\rg\ot |\ep\rg) = 	\sum_{l=0}^3(\sg_l
|\alpha\rg)\ot |e'_l\rg,
\label{e5.6}
\end{equation}
where the $|e'_l\rg$ are obtained from the $|e_l\rg$ by an anti-linear
operation (``time reversal'') as explained below.

	Choose a reference orthonormal basis $\{|r_j\rg\}$ for $\E$, and for
any $|e\rg$ in $\E$, define its ``complex conjugate'' $|e^*\rg$ as the vector
with components
\begin{equation}
  \lg r_j|e^*\rg = \lg r_j|e\rg^*.
\label{e5.7}
\end{equation}
Likewise, if $F$ is any operator on $\E$, the operator $F^*$ is defined by the
matrix elements
\begin{equation}
  \lg r_j|F^*|r_k\rg = \lg r_j|F|r_k\rg^*,
\label{e5.8}
\end{equation}
that is, the matrix of $F^*$ in this basis is the complex conjugate of that of
$F$.  Of course, both $|e^*\rg$ and $F^*$ depend upon which reference basis is
used, but that is immaterial as long as the latter is held fixed throughout the
following argument.  For later reference, note the relationships
\begin{equation}
  \lg e^*|f^*\rg= \lg e|f\rg^*,\quad \Tr[F^*] = (\Tr[F])^*.
\label{e5.9}
\end{equation}

	Now define the $|e'_l\rg$ introduced in (\ref{e5.6}) by
\begin{equation}
  |e'_0\rg = |e^*_0\rg; \quad |e'_q\rg = - |e^*_q\rg \hbox{ for } q=1,2,3.
\label{e5.10}
\end{equation}
It is easy to check that (\ref{e2.8}) and (\ref{e2.9}) are satisfied by the
matrix elements
\begin{equation}
  E'_{lm} = \lg e'_l|e'_m\rg
\label{e5.11}
\end{equation}
if they are satisfied by the $E_{lm}$, so $V'$ with the $|e'_l\rg$ defined in
(\ref{e5.10}) is, indeed, an isometry.  The corresponding superoperator $\hat
V'$, see (\ref{e2.11}) to (\ref{e2.13}), can be written as
\begin{equation}
  \hat V'(\sg_l) = \sum_m \sg_m\ot \hat F'_{lm}
\label{e5.12}
\end{equation}
with
\begin{equation}
  \hat F'_{lm} = \sum_{jk} L(jk;lm) |e'_j\rg\lg e'_k|.
\label{e5.13}
\end{equation}

	It is then a straightforward exercise to show, using the properties of
$L(jk;lm)$ given in App.~A, that
\begin{equation}
  \hat F'_{lm} = \pm \hat F^*_{lm},
\label{e5.14}
\end{equation}
where the sign is $-$ for $l=0,m>0$ and for $l>0,m=0$; and is $+$ in all other
cases.  Consequently, the elements of the $B'$ matrix corresponding to $V'$,
see (\ref{e2.19}), are given by
\begin{equation}
  B'_{lm}= \Tr[\hat F'_{lm}] = \pm B_{lm}
\label{e5.15}
\end{equation}
with the sign following the same rule as in (\ref{e5.14}); note that the
complex conjugate can be ignored when we use the trace formula in (\ref{e5.9}),
because the $B_{lm}$ are real.  Therefore $B'$ is the same as $B$, except that
\begin{equation}
  B'_{0q} = -B_{0q} \hbox{ for } q=1,2,3;
\label{e5.16}
\end{equation}
we need not concern ourselves with the $B_{q0}$, because they vanish. That is,
$B'$ is obtained from $B$ by reversing the sign of $\delbb$, precisely what we
need in order to investigate (\ref{e5.1}).

	The final step in proving (\ref{e5.1}) is to use (\ref{e3.13}) for both
$V'$ and $V$ to show that
\begin{equation}
  \Omega'_\E(\mhb) = \sum_{q=1}^3 m_q\hat F'_{0q} 	= - \sum_{q=1}^3 m_q
\hat F^*_{0q} = -\Omega^*_\E(\mhb).
\label{e5.17}
\end{equation}
Since $\Omega_\E(\mhb)$ is a Hermitian operator, its eigenvalues are the same
as those of $\Omega^*_\E(\mhb)$, and since the quality factor is the sum of the
absolute values of the eigenvalues, we conclude that
\begin{equation}
  \Q'_\E(\mhb) = \Tr(|\Omega'_\E(\mhb)|) = 	\Tr(|\Omega_\E(\mhb)|)
=\Q_\E(\mhb),
\label{e5.18}
\end{equation}
which is (\ref{e5.1}).

	\subsection{Proof of the Concavity Property}
\label{s5c}

	To establish (\ref{e5.2}) we proceed as follows.  Let $V^1$ and $V^2$
be any two isometries of the type (\ref{e2.4}):
\begin{equation}
  r=1,2:\quad V^r|\alpha\rg = \sum_{l=0}^3 (\sg_l |\alpha\rg)\ot |e_l^r\rg.
\label{e5.19}
\end{equation}
Since there is no constraint upon the dimension of the space $\E$, we can
assume that the vectors $|e_l^r\rg$ are chosen in such a way that
\begin{equation}
  \lg e_l^1 | e_m^2\rg =0
\label{e5.20}
\end{equation}
for all $l$ and $m$; in other words, the subspace of $\E$ containing the
vectors $\{|e_l^1\rg\}$ is orthogonal to that containing the vectors
$\{|e_l^2\rg\}$.  This, of course, places no restriction upon the matrices
\begin{equation}
  E_{lm}^r = \lg e_l^r | e_m^r\rg,
\label{e5.21}
\end{equation}
aside from the isometry conditions (\ref{e2.8}) and (\ref{e2.9}), which must be
satisfied separately for $r=1$ and $2$.

	Now construct a third isometry
\begin{equation}
  V=\sqrt{p_1}V^1 + \sqrt{p_2}V^2,
\label{e5.22}
\end{equation}
where $p_1$ and $p_2$ are non-negative numbers whose sum is $1$.  That this $V$
is an isometry satisfying (\ref{e2.7}) is a consequence of the fact that both
$V^1$ and $V^2$ are isometries, along with (\ref{e5.20}), which implies that
$(V^1)\ad V^2$ and $(V^2)\ad V^1$ both vanish.  The superoperator corresponding
to (\ref{e5.22}) takes the form
\begin{equation}
  \Vh(\kp) = V\kp V\ad = p_1 \Vh^1 (\kp)+p_2 \Vh^2 (\kp) +\sqrt{p_1p_2}\, \hat
W(\kp),
\label{e5.23}
\end{equation}
where $\Vh^r$ is the superoperator corresponding to $V^r$, and
\begin{equation}
  \hat W(\kp) = V^1\kp(V^2)\ad + V^2\kp(V^1)\ad
\label{e5.24}
\end{equation}
has the property that
\begin{equation}
  \Tr_\E [\hat W(\kp)]=0.
\label{e5.25}
\end{equation}
To see that (\ref{e5.25}) is the case, note that (\ref{e5.20}) implies that we
can choose an orthonormal basis $\{|\ep_k^r\rg\}$ for $\E$, with $r=1$ or $2$,
and $k$ taking on as many values as necessary, such that
\begin{equation}
  \lg e_l^1 | \ep_k^2\rg =0 = \lg e_l^2 | \ep_k^1\rg
\label{e5.26}
\end{equation}
whatever the values of $l$ and $k$.  Evaluating the trace (\ref{e5.25}) in this
basis shows that it is zero.

	Hence the partial trace of (\ref{e5.23}) over $\E$ yields
\begin{equation}
  \Vh_\B(\kp) = p_1\Vh_\B^1(\kp)+ p_2\Vh_\B^2(\kp),
\label{e5.27}
\end{equation}
and, as a consequence of (\ref{e2.18}), the corresponding $4\times4$ $B$
matrices of Sec.~\ref{s2} are related by
\begin{equation}
  B=p_1B^1 + p_2 B^2.
\label{e5.28}
\end{equation}
Since $B^1$ and $B^2$ can be any two $B$ matrices which are physically
possible---the isometries $V^1$ and $V^2$ are arbitrary---and $p_1$ and $p_2$
any positive numbers whose sum is one, (\ref{e5.28}) shows that the set of
physically possible $B$ matrices generated by isometries of the type
(\ref{e2.4}) is convex.

	Tracing (\ref{e5.23}) over $\B$ yields
\begin{equation}
  \Vh_\E(\kp) = p_1 \Vh_\E^1 (\kp)+p_2 \Vh_\E^2 (\kp) 	+\sqrt{p_1p_2}\, \hat
W_\E(\kp),
\label{e5.29}
\end{equation}
and $\hat W_\E(\kp)$ does not (in general) vanish. If we write the identity on
$\E$ in the form
\begin{equation}
  I_\E = P^1 + P^2
\label{e5.30}
\end{equation}
where
\begin{equation}
  P^r= \sum_j |\ep_j^r\rg\lg\ep_j^r|
\label{e5.31}
\end{equation}
is a projector onto the subspace containing the $\{|e_l^r\rg\}$, then
\begin{equation}
  P^1\Vh_\E(\kp) P^1 = p_1 \Vh_\E^1 (\kp),\quad 	P^2\Vh_\E(\kp) P^2 =
p_2 \Vh_\E^2 (\kp),
\label{e5.32}
\end{equation}
and theorem 2 in App.~B tell us that
\begin{equation}
   \Tr(|\Vh_\E(\kp)|) \geq p_1\Tr(|\Vh_\E^1(\kp)|)+ p_2\Tr(|\Vh_\E^2(\kp)|).
\label{e5.33}
\end{equation}
Substituting $\kp = \half\bold{\hat m\cdot\sg}$ into this expression yields
(\ref{e5.2}).

		\section{Optimal Pairs and Optimization Map}
\label{s6}
	\subsection{General Conditions}
\label{s6a}

	In Sec.~\ref{s4} we showed that if $\bb$ is the three vector of
principal axes for the ellipsoid of the $\B$ channel, there is a largest
ellipsoid for the $\C$ channel, characterized by principal axes $\cb$ and given
by the function $G(\bb)$, (\ref{e4.1}), which represents the best possible copy
in $\C$ if $\bb$ is held fixed.  Note that the components of $\cb=G(\bb)$ are
all positive, whereas those of $\bb$ may be either positive or negative.  In
this section we examine the optimization function $G(\bb)$ in a bit more
detail, and deduce necessary and sufficient conditions for an optimal {\it
pair} $(\bb,\cb)$ of copies.

	If $\fb=(f_1,f_2,f_3)$ is a three vector with real components, we
define its absolute value to be
\begin{equation}
  \fb^+=(|f_1|,|f_2|,|f_3|).
\label{e6.1}
\end{equation}
We shall write $\fb\geq 0$ provided each $f_p$, $1\leq p\leq 3$, is
non-negative, and $\fb >0$ if, in addition, at least one $f_p$ is strictly
positive.  The relationship $\fb\geq \gb$ means $\fb-\gb \geq 0$, or $f_p\geq
g_p$ for $1\leq p \leq 3$, whereas $\fb > \gb$ means $\fb-\gb > 0.$ The same
notation will be used for a four vector such as
$\bt=(\bt_0,\bt_1,\bt_2,\bt_3)$; $\bt^+$ is its absolute value in the sense of
(\ref{e6.1}), and $\bt > 0$ means that at least one component is strictly
positive and the rest are non-negative.

	A three vector $\bb$ will be said to be {\it possible} provided it
satisfies (\ref{e4.10}) and is thus a possible collection of semi-axes (some of
which may be negative) for the ellipsoid in $\B$, while a {\it possible pair}
$(\bb,\cb)$ is one that corresponds to some conceivable copying machine, i.e.,
is allowed by the laws of quantum mechanics.  The pair $(\bbr,\cbr)$ will be
said to be {\it better} (in quality) than $(\bb,\cb)$ provided
\begin{equation}
  \bb{\vphantom{\bbr}}^+\leq\bbr^+,\quad \cb^+\leq\cbr^+,
\label{e6.2}
\end{equation}
and at least one of these inequalities is strict, e.g., $\cb^+ < \cbr^+$.  If
both inequalities are equalities, the pairs are {\it equivalent} (in quality).
We use absolute values in (\ref{e6.2}) because they alone enter the quality
function, see (\ref{e3.12}).
	Finally, an {\it optimal pair} $(\bb,\cb)$ is one for which no other
{\it possible} pair can be found which is better.  Note that if $(\bb,\cb)$ is
optimal, so also is $(\cb,\bb)$, for it obviously does not matter which output
channel is labeled $\B$ and which is labeled $\C$.

	For any possible $\bb$, the argument in Sec.~\ref{s4}, supplemented by
the results in Sec.~\ref{s5}, tells us that $(\bb,G(\bb))$ is a possible pair,
and for any other possible pair of the form $(\bb,\cbr)$ it is the case that
\begin{equation}
  \cbr^+ \leq G(\bb);
\label{e6.3}
\end{equation}
recall that, by construction, $G(\bb)\geq 0$, so it is not necessary to take
the absolute value on the right side.  This does {\it not} mean that given any
possible $\bb$, $(\bb,G(\bb))$ is an optimal pair in the sense defined above,
because there might very well be a positive optimal pair $(\bb',\cb')$ with
$\bb^+ < \bb'$ and $G(\bb)\leq \cb'$.  Indeed, this can occur even in cases in
which $\bb=\bb^+$ has all positive components.  However, if $(\bb,\cb)$ is an
optimal pair, then so is $(\bb,G(\bb))$, and
\begin{equation}
  \cb^+=G(\bb).
\label{e6.4}
\end{equation}
The reason is that the pair $(\bb,G(\bb))$ is possible, and were it better than
$(\bb,\cb)$, that is, were (\ref{e6.4}) the inequality allowed by (\ref{e6.3}),
then $(\bb,\cb)$ would not be an optimal pair.  Thus we see that if $(\bb,\cb)$
is optimal, so is $(\bb,\cb^+)$, and, by the same sort of argument,
$(\bb^+,\cb)$ and $(\bb^+,\cb^+)$ are also optimal pairs.

	An optimal pair $(\bb,\cb)$ with $\bb\geq 0$ and $\cb\geq 0$ will be
called a {\it positive optimal pair}.  Note that if we reverse the signs of
some of the components of $\bb$ or $\cb$, or both, the resulting pair will also
be optimal {\it provided it is possible}.  Thus the task of characterizing
optimal pairs can be divided into two parts: finding necessary and sufficient
conditions for positive optimal pairs, and conditions under which reversing
some of the signs in a positive optimal pair results in a possible, and
therefore optimal pair.

	The optimization map $G(\bb)$ described in Sec.~\ref{s4} is obtained
through a series of steps which can be represented schematically in the form:
\begin{equation}
  b \llra \bt^2 \lla \bt \llra \gm \lra \gm^2 \llra c
\label{e6.5}
\end{equation}
in terms of the four-vectors introduced in Sec.~\ref{s4}; see, e.g.,
(\ref{e4.5}) and (\ref{e4.16}). Here $\llra$ stands for the reversible
transformation obtained by multiplying a four vector by the matrix $\Lm$ and
perhaps a positive constant: see the explicit expressions in (\ref{e4.6}),
(\ref{e4.14}), and (\ref{e4.16}).  The single arrow means taking the square of
each component; thus $\bt^2$ is $(\bt_0^2,\bt_1^2,\bt_2^2,\bt_3^2)$.  This
operation is not reversible because it is many to one or, equivalently, because
the square root can be either positive or negative.  (Note that $\bt^2 > 0$ is
a consequence of the positivity of the matrix $E_{lm}$, see (\ref{e2.6}) and
(\ref{e4.2}), so we do not have to worry about imaginary roots.)  Thus in going
from the left side to the right side of (\ref{e6.5}) we have to specify the
square root of $\bt^2$.  The map $G(\bb)$ defined in Sec.~\ref{s4} employs the
positive root, $\bt >0$.

	As a consequence, whereas one or more of the components of $\bb$ may be
negative (recall that $b_0=1=c_0$), the components of $\cb$, which can be
expressed in terms of those of $\bt$ using (\ref{e4.17}), are always
non-negative: $G(\bb) \geq 0$. On the other hand, $\bt > 0$ does not imply that
the components of $\gm$ in (\ref{e6.5}) are positive; indeed, the counterpart
of (\ref{e4.17}),
\begin{equation}
      b_q = 2(\gm_0\gm_q+\gm_{q'}\gm_{q''})
\label{e6.6}
\end{equation}
for $1\leq q \leq 3$, shows that if some $b_q$ is negative, some of the $\gm_j$
must be negative as well.

	If in (\ref{e6.6}) we replace every component of $\gm$ by its absolute
value, the result will be a vector $\bb'$ with, obviously,
\begin{equation}
  \bb^+\leq\bb' = G(\cb),
\label{e6.7}
\end{equation}
since $G(\cb)$ results from following the chain in (\ref{e6.5}) from right to
left with $\gm$ replaced by the positive square root of $\gm^2$.

	By examining (\ref{e6.6}) for each $p$ in turn, and noting that $\gm_0$
must be strictly positive (half the sum of the $\bt_j$), it is easy to show
that a necessary and sufficient condition for equality between $\bb^+$ and
$\bb'$ in (\ref{e6.7}) is that
\begin{equation}
  \gm_4 := \gm_0\gm_1\gm_2\gm_3 \geq 0.
\label{e6.8}
\end{equation}
Consequently, if $\gm_4$ is negative, $(\bb,\cb)$ with $\cb=G(\bb)$ cannot be
an optimal pair, since $(\bb',\cb)$ with $\bb'=G(\cb)$ is possible, and better.
In addition,
\begin{eqnarray}
  &&b_1b_2b_3/8 = \gm_0^2\gm_1^2\gm_2^2 + \gm_0^2\gm_1^2\gm_3^2
+\gm_0^2\gm_2^2\gm_3^2
\nonumber\\
&&\quad + \gm_1^2\gm_2^2\gm_3^2 + \gm_4 (\gm_0^2 + \gm_1^2 +\gm_2^2 +\gm_3^2),
\label{e6.9}
\end{eqnarray}
a consequence of (\ref{e6.6}), tells us that if $(\bb,\cb)$ is an optimal pair,
then
\begin{equation}
  b_1b_2b_3 \geq 0.
\label{e6.10}
\end{equation}
The reason is that if (\ref{e6.10}) is violated, so is (\ref{e6.8}), which
would mean that (\ref{e6.7}) is a strict inequality, and $(G(\cb),\cb)$ is
better than $(\bb,\cb)$.  But this is not possible if $(\bb,\cb)$ is optimal.
Note that (\ref{e6.8}) is stronger than (\ref{e6.10}), so even if (\ref{e6.10})
is satisfied, $(\bb,G(\bb))$ need not be optimal.  Of course, (\ref{e6.10})
with $b$ replaced by $c$ must also hold for any optimal pair $(\bb,\cb)$.

	However, (\ref{e6.10}) is enough to settle the question of whether,
given a {\it positive} optimal pair $(\bb,\cb)$, a pair $(\bb',\cb')$ obtained
by reversing the signs of one or more components of $\bb$ or $\cb$ (or both) is
possible, and hence optimal.  The answer is that $\bb'$ is either $\bb$ itself,
or obtained from $\bb$ by reversing the signs of {\it precisely two} components
of $\bb$, leaving the sign of the other component fixed.  If $\bb$ has a
component equal to zero, this can, but need not be counted among the two whose
sign is reversed, and thus any or all of the non-zero components can be changed
in sign without violating (\ref{e6.10}) or the two component rule just
mentioned. The same relationship must hold between $\cb'$ and $\cb$.  Now since
reversing the sign of two of the components of $\bb$ can be achieved using a
suitable unitary transformation on $\B$, see the discussion in the paragraph
preceding (\ref{e2.28}), and a similar comment applies to $\cb$, a
$(\bb',\cb')$ obtained in this way is surely possible, and therefore optimal.
On the other hand, reversing the sign of just one component of $\bb$, assuming
all three components are non-zero, will violate (\ref{e6.10}), and thus the
resulting pair is not possible, since if it were possible it would be optimal.
And the same comment applies to $\cb$ when its three components are all
positive.

	In summary, the task of identifying all optimal pairs $(\bb,\cb)$
reduces to that of identifying all positive optimal pairs.  All others are
obtained from these by changing the signs of two components of $\bb$ and/or two
components of $\cb$, since this operation always yields an optimal pair, and
any optimal pair which is non-positive is the result of such an operation.  In
addition, if $(\bb,\cb)$ is a positive optimal pair, it follows from the
definition of optimality and the properties of the map $G$, see (\ref{e6.4}),
that
\begin{equation}
  \cb=G(\bb),\quad \bb=G(\cb),
\label{e6.11}
\end{equation}
that is, the chain (\ref{e6.5}) will work in either direction, and, therefore,
\begin{equation}
  \bt > 0,\quad \gm =\half\Lm\cdot\bt > 0.
\label{e6.12}
\end{equation}

	However (\ref{e6.11}), or the equivalent (\ref{e6.12}), while it
implies (\ref{e6.8}) and the corresponding condition for $\bt$, and is thus
stronger than (\ref{e6.10}) or the corresponding condition on $\cb$, is a
necessary, but not sufficient condition for $(\bb,\cb)$ to be a positive
optimal pair, because (as noted following (\ref{e6.3})) there might be a pair
$(\bb',\cb')$ for which $ \cb'=G(\bb')$ and $\bb'=G(\cb')$, with $\bb <
\bb'$ and $\cb < \cb'$.  To examine this possibility, it is useful to study the
Jacobian matrix of the map $G$.  As shown in App.~\ref{ac}, one can relate the
differentials of $\bb$ and $\cb$ through
\begin{equation}
  d\cb = (16 \bt_4)^{-1} \Jb\cdotb\, d\bb 
\label{e6.13}
\end{equation}
or its inverse
\begin{equation}
  d\bb=(16 \gm_4)^{-1} \Kb\cdotb\, d\cb,
\label{e6.14}
\end{equation}
where $\gm_4$ is defined in (\ref{e6.8}), and $\bt_4$ is the corresponding
quantity for $\bt$.  The $3\times3$ matrix $\bold{J}$ has components
\begin{equation}
  J_{qq} = h_{q'} h_{q''}, \quad J_{qq'} = -h_{q'}c_{q''},\quad J_{q'q} = -h_q
c_{q''},
\label{e6.15}
\end{equation}
while those of $\Kb$ are
\begin{equation}
  K_{qq} = h_{q'} h_{q''}, \quad K_{qq'} = -h_{q'}b_{q''},\quad K_{q'q} = -h_q
b_{q''},
\label{e6.16}
\end{equation}
and $\hb=(h_1,h_2,h_3)$ is a vector whose components are defined by
\begin{equation}
  h_q=2(\bt_0\bt_q -\bt_{q'}\bt_{q''}) = 2(\gm_0\gm_q -\gm_{q'}\gm_{q''}).
\label{e6.17}
\end{equation} 

	Let us assume that $\gm_4 > 0$, as will be the case except on the
boundaries of the region (\ref{e6.12}); note that this implies, (\ref{e6.6}),
that each component of $\bb$ is strictly positive. If we can find a $d\cb$
whose components are all positive and which, inserted in (\ref{e6.13}), yields
a $d\bb$ whose components are also positive, then it is clear that $(\bb,\cb)$
cannot be a positive optimal pair, for there is a better $(\bb',\cb')$ in its
immediate vicinity.  It turns out, see App.~\ref{ad}, that under the conditions
(\ref{e6.12}), at most one of the $h_q$ can be negative.  If we suppose that
\begin{equation}
  h_1 < 0,\quad h_2 > 0,\quad h_3 > 0,
\label{e6.18}
\end{equation}
 the elements $K_{q1}$ of the first column of the $\Kb$ matrix are all strictly
positive, and therefore if $d\cb=(1,0,0)$ is inserted in (\ref{e6.14}), the
resulting $d\bb$ will have all three components strictly positive.  Clearly the
same will be true for a $d\cb$ with all positive components which is
sufficiently close to the one we have just considered, and, consequently,
(\ref{e6.18}), or any other case in which two components of $\hb$ are positive
and one is negative, is inconsistent with $(\bb,\cb)$ being a positive optimal
pair, as long as $\gm_4 > 0$.

	What this suggests is that in addition to (\ref{e6.11}) or
(\ref{e6.12}), we need the condition
\begin{equation}
  \hb \geq 0,
\label{e6.19}
\end{equation}
in order to ensure that $(\bb,\cb)$ is a positive optimal pair.
	Note that we have not established with complete mathematical rigor that
(\ref{e6.19}) is a necessary condition for a positive optimal pair, since the
argument given above no longer works (at least in any simple form) if, for
example, $h_2 > 0$ in (\ref{e6.18}) is replaced by $h_2=0$.  Such ``boundary
cases'' require additional study, which we have carried through in some simple
cases, but not in complete detail.  In addition, even if (\ref{e6.19}) is
necessary, it does not follow that it is sufficient, in conjunction with
(\ref{e6.11}), to guarantee that $(\bb,\cb)$ is a positive optimal pair,
because it refers only to local properties of the $G$ map.  While the region
defined by (\ref{e6.19}) along with (\ref{e6.12}) is connected, we know of no
obvious reason why a given $(\bb,\cb)$ in this region must be connected to a
better $(\bbr,\cbr)$, assuming one exists, through some path along which each
component changes monotonically.  Such a path can be ruled out by a minor
extension of the above analysis as long as everywhere on it all the components
of $\hb$, together with $\gm_4$ are strictly positive. But paths along which
some components of $\bb$ or $\cb$ increase while others decrease, followed
later by increases and decreases of other components, the final result being an
increase in all components, are easy to imagine.

	Having not succeeded in supplying an analytical argument, we attacked
the problem numerically in the following way.  Vectors $\bb$ were chosen at
random in the ``good'' region defined by (\ref{e6.12}) and (\ref{e6.19}), and,
for each such $\bb$, further vectors $\bbr$ lying in the ``good'' region and
also satisfying $\bbr > \bb$ were chosen at random, and we checked to see
whether
\begin{equation}
  G(\bbr) \geq G(\bb).
\label{e6.20}
\end{equation}
With some 4000 choices of $\bb$, and for each of these 100,000 choices of
$\bbr$, we found no cases in which (\ref{e6.20}) was satisfied, whereas if we
made a similar test using $\bb$ vectors lying outside the ``good'' region, a
random search turned up many examples of vectors $\bbr > \bb$ satisfying
(\ref{e6.20}).

	Consequently, we believe that the necessary and sufficient condition
for $(\bb,\cb)$ to be a positive optimal pair is that $\bb$ and $\cb$ are both
generated using (\ref{e6.5}) from a four vector $\bt$ satisfying the conditions
\begin{equation}
  0\leq \bt_q \leq \bt_0,\quad \bt_0\bt_q \geq \bt_{q'}\bt_{q''}
\label{e6.21}
\end{equation}
for $q=1,2$ and $3$---note that the second inequality is equivalent to
(\ref{e6.19})---along with the normalization condition (\ref{e4.8}).

	Let us define the {\it class} $P$ to be the collection of all
four-vectors which satisfy the two conditions in (\ref{e6.21}) for all $q$
between 1 and 3, whatever the value of $\sum_j\bt_j^2$.  Also, we shall say
that any two vectors $\xi$ and $\eta$ belonging to class $P$ are in (or have)
the {\it same order} provided, for $1\leq p\leq 3$ and $1\leq q\leq 3$,
\begin{equation}
  \sign(\xi_p-\xi_q)=\sign(\eta_p-\eta_q),
\label{e6.22}
\end{equation}
where $\sign(x)$ is $+1$, $0$ or $-1$ for $x$ positive, zero, or negative.  For
example, if $\xi_1\leq\xi_2\leq\xi_3$, then $\eta_1\leq\eta_2\leq\eta_3$, and
vice versa; $\xi_1=\xi_2$ if and only if $\eta_1=\eta_2$.  These definitions
lead to the following useful result:

	{\bf Theorem.} If $\xi$ is a four vector in class $P$, and $\eta$
another four vector obtained from $\xi$ by one of the three operations (a),
(b), and (c) given below, or by the application of any combination of these
operations carried out in succession, then $\eta$ is also in class $P$, and
$\eta$ has the same order as $\xi$.  The three operations are:

	(a) $\eta = k\xi = (k\xi_0,k\xi_1,k\xi_2,k\xi_3)$ for a constant $k >
0$.

	(b) $\eta=\xi^a = (\xi_0^a,\xi_1^a,\xi_2^a,\xi_3^a)$ for an exponent
$a>0$.

	(c) $\eta = \Lm\cdot\xi$.

	The proof is straightforward, except for (c), for which see
App.~\ref{ad}.

	The theorem is useful when applied to (\ref{e6.5}), because each step
on the way from $b$ to $c$ is one of the operations (a), (b) or (c), or a
combination of (a) and (c).  Consequently, if $\beta$ satisfies the conditions
(\ref{e6.21}), they are also satisfied by $\gm$, and also by $b$ and $c$.  In
particular, if $\bt$ is normalized,
\begin{equation}
  \sum_{j=0}^3 \bt_j^2 = 1,
\label{e6.23}
\end{equation}
then for $1\leq q\leq 3$,
\begin{equation}
  0\leq b_q\leq b_0=1, \quad b_q \geq b_{q'} b_{q''}.
\label{e6.24}
\end{equation}
Note that (\ref{e6.24}) is not only implied by (\ref{e6.21}) together with the
normalization (\ref{e6.23}), it also implies (\ref{e6.21}) and (\ref{e6.23}),
and also the set of conditions for $c=(1,\cb)$ which correspond to
(\ref{e6.24}).

	Indeed, (\ref{e6.24}) is a {\it necessary and sufficient} condition for
a triple of numbers $\bb=(b_1,b_2,b_3)$ to belong to a positive optimal pair,
assuming that the latter can be characterized by (\ref{e6.12}) and
(\ref{e6.19}).  The reason is that (\ref{e6.24}) tells us that $b=(1,\bb)$
belongs to class $P$, and therefore the four vector $\beta^2$ belongs to class
$P$ and has non-negative components.  This, together with the normalization
(\ref{e6.23}) corresponding to $b_0=1$ in (\ref{e6.24}) means that $\bb$ is
physically possible.  Hence the pair $(\bb,G(\bb))$ is also possible, it is
obviously positive, and it is optimal, since (\ref{e6.21}), thus (\ref{e6.19}),
is satisfied.

	In addition, the theorem tells us that if (\ref{e6.24}) holds and the
semi-axes are in the order
\begin{equation}
  b_1 \leq b_2 \leq b_3,
\label{e6.25}
\end{equation}
then the $\bt_q$, $\gm_q$ are in the same order, and likewise
\begin{equation}
  c_1 \leq c_2 \leq c_3,
\label{e6.26}
\end{equation}
for $\cb = G(\bb)$, the other member of the optimal pair.  In addition, if two
of the $\B$ semi-axes are equal to each other, say $b_1=b_2$, then the
corresponding $\C$ semi-axes are also equal, $c_1=c_2$.  Note, however, that
there is no reason to expect $b_1$ to be equal to $c_1$; one ellipsoid can be
small and the other large.

	In summary, a necessary and sufficient condition that $(\bb,\cb)$ be a
positive optimal pair is that one member of the pair satisfy (\ref{e6.24}) and
the other member be obtained by applying the optimization map $G$ to the first.
The second member will then also satisfy the conditions (\ref{e6.24}), and
$\bb$ and $\cb$ will have the same order.  All optimal pairs which are not
positive optimal pairs are obtained from some positive optimal pair by
reversing the signs of two components of the first member of the pair, and/or
two components of the second member.

	\subsection{Particular Examples}
\label{s6b}

\begin{figure}
\epsfxsize=3in
\epsfbox{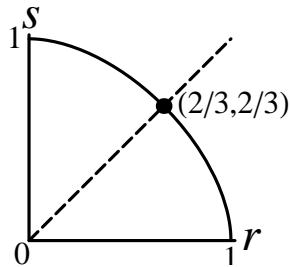}
\caption{The relationship between the radii ($s$ and $r$) of the two output Bloch spheres in
the case of isotropic copying. Identical copies occur when $r=s=2/3$.}
\label{f1}
\end{figure}

	Isotropic copies in which
\begin{equation}
  b_1=b_2=b_3= r,\quad c_1 = c_2 = c_3=s,
\label{e6.27}
\end{equation}
that is, the Bloch ellipsoids for $\B$ and $\C$ are spheres of radii $r$ and
$s$, are obtained by letting
\begin{equation}
  \beta_1=\beta_2=\beta_3=\sqrt{(1-\beta_0^2)/3}.
\label{e6.28}
\end{equation}
Using (\ref{e4.9}) and (\ref{e4.17}), one can show that
\begin{equation}
  s=\half\left[1-r+\sqrt{(1-r)(1+3r)} \right],
\label{e6.29}
\end{equation}
and, by symmetry, $r$ as a function of $s$ is given by precisely the same
functional form.  The relationship between the two is shown in Fig.~\ref{f1}.
Since $r$ is between 0 and 1, the condition (\ref{e6.24}) for optimal pairs is
satisfied.  In other words, any point that lies outside the region enclosed by
this curve and the two axes is prohibited; this is identical to the no-cloning
bound in \cite{r12}.

	As one might expect, there is a certain ``complementarity'' between the
quality of the copies emerging in the $\B$ and $\C$ channels: as $r$ increases,
$s$ decreases.  Additional insight into the source of this complementarity
comes from studying anisotropic situations.  Consider, in particular, the
extremely anisotropic case in which
\begin{equation}
  \beta_1=\beta_2 = 0, \quad 0 \leq \beta_3 \leq 1/\sqrt{2};
\label{e6.30}
\end{equation}
the upper bound on $\beta_3$ ensures that (\ref{e6.21}) is satisfied.  The
corresponding $\bb$ and $\cb$ for optimal copies are given by
\begin{eqnarray}
  &&b_1=b_2=1-2\beta_3^2,\quad b_3=1,
\label{e6.31}\\
  &&c_1=c_2=0,\quad c_3=2\beta_3\sqrt{1-\beta_3^2}.
\label{e6.32}
\end{eqnarray}
Note that as $\beta_3$ increases, the quality factor $c_3$ for the third
principal mode of the $\C$ copy increases whereas the factor for the same mode
of $\B$ remains perfect, $b_3=1$.  On the other hand, the quality factors for
modes 1 and 2 of $\B$ decrease as $\beta_3$ increases.  	 Thus improving
the quality of a particular mode in one channel leads to a decrease in optimal
quality of all the ``perpendicular'' modes of the other channel.  In
particular, $c_1$ and $c_2$ are 0 (worthless copies) in this case because of
the perfect quality for mode 3 of $\B$.

		\section{Quantum Circuit}
\label{s7}

A complicated unitary transformation can often be decomposed into several
simpler transformations and expressed in a pictorial fashion by using a quantum
circuit.  This is helpful both for understanding the transformation and for
implementing it in a systematic way.  It is worth noting that any unitary
transformation can be produced by a circuit which only uses two-qubit XOR gates
along with an appropriate collection of one-qubit gates \cite{r17}.
	
\begin{figure}
\epsfxsize=5in
\epsfbox{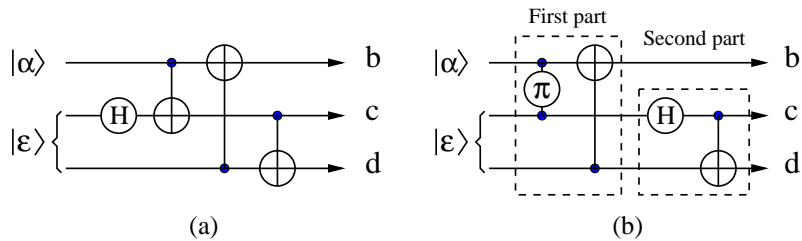}
\caption{(a) Optimal copy circuit.  The qubit to be copied enters the circuit
as $|\alpha\rg$ on the left, and the copies emerge in $b$ and $c$ on the right.
(b) Alternative circuit, see discussion in text.}
\label{f2}
\end{figure}

From Sec.~\ref{s4}, the optimal copying procedure corresponding to $\cb=G(\bb)$
can always be achieved, for any given $\bb$, by using a transformation which
generates the isometry (\ref{e4.13}).  One example is given by the circuit in
Fig.~\ref{f2}(a), where the input state is a tensor product
$|\alpha\rg\ot|\ep\rg$, with $|\alpha\rg$ arbitrary, and
\begin{equation}
  |\ep\rg = \beta_0|00\rg +\beta_1|01\rg +\beta_2|11\rg +\beta_3|10\rg.
\label{e7.1}
\end{equation}
Since $|\ep\rg$ involves only real coefficients, it can be produced by a
preparation circuit as shown in \cite{r10}.  The three horizontal lines
represent the three qubits going from left to right, as time increases, through
four quantum gates.  The order of the qubits from top to bottom is the same as
the left-to-right order in (\ref{e4.13}).

	The first gate in Fig.~\ref{f2}(a) produces the Hadamard transform
\begin{eqnarray}
 &&|0\rg\mt |\!+\!\rg=(|0\rg+|1\rg)/\sqrt{2},\nonumber\\ &&|1\rg\mt
|\!-\!\rg=(|0\rg-|1\rg)/\sqrt{2},
\label{e7.2}
\end{eqnarray}
on a single qubit.  The remaining three gates are XOR gates acting on different
pairs of qubits; each corresponds to the transformation
\begin{eqnarray}
|00\rg\mt|00\rg,&\quad&|10\rg\mt|11\rg,\nonumber\\
|01\rg\mt|01\rg,&\quad&|11\rg\mt|10\rg,
\label{e7.3}
\end{eqnarray}
in which the left qubit in this formula is the ``controlling'' and the right
qubit the ``controlled'' qubit.  If the controlling qubit is $|0\rg$, the
controlled qubit is left unchanged; if the controlling qubit is $|1\rg$, the
controlled qubit is flipped from $|0\rg$ to $|1\rg$, or $|1\rg$ to $|0\rg$,
which is known as ``amplitude-flipping''.  In Fig.~\ref{f2}(a) the controlling
qubit is indicated by a solid dot, and the corresponding controlled qubit by a
plus inside a circle.  Thus the controlling qubit is the top, bottom, and
middle qubit in the first, second, and third XOR gates.

	The transformation produced by the circuit in Fig.~\ref{f2}(a) can be
more easily understood if one uses the alternative circuit in Fig.~\ref{f2}(b),
which produces exactly the same overall unitary transformation.  It employs
another gate, indicated by $\pi$ inside a circle connecting two black dots,
which produces the transformation
\begin{eqnarray}
|00\rg\mt|00\rg,&\quad&|10\rg\mt|10\rg,\nonumber\\
|01\rg\mt|01\rg,&\quad&|11\rg\mt-|11\rg,
\label{e7.4}
\end{eqnarray}
where the phase of $|11\rg$ is multiplied by a factor of $e^{i\pi}=-1$.  It
flips one qubit between $|\!+\!\rg$ and $|\!-\!\rg$, see (\ref{e7.2}), if the
other qubit is $|1\rg$. This is known as a ``phase flip'', in contrast to the
``amplitude flip'' produced by an XOR gate.  Note that in this case either
qubit can be regarded as controlling the other.

	Both circuits in Fig.~\ref{f2} carry out the transformation
\begin{equation}
|\alpha\rg\ot |\ep\rg\mt
\sum_l(\sg_l |\alpha\rg)\ot\bt_l|\eh_l\rg
\label{e7.5}
\end{equation}
corresponding to (\ref{e2.4}) combined with (\ref{e4.3}), and thus the isometry
(\ref{e4.13}) on the input qubit $|\alpha\rg$.  But Fig.~\ref{f2}(b) is a bit
easier to interpret.  In the {\it first part} of the circuit, the lower two
qubits control the top qubit: the middle qubit, if it is $|1\rg$, flips the
phase, and the bottom qubit, if $|1\rg$, flips the amplitude.  Thus this part
of the circuit results in
\begin{eqnarray}
|\alpha\rg\ot|00\rg&\mt&(\sg_0|\alpha\rg)\ot|00\rg,\nonumber\\
|\alpha\rg\ot|01\rg&\mt&(\sg_1|\alpha\rg)\ot|01\rg,\nonumber\\
|\alpha\rg\ot|11\rg&\mt&(\sg_1\sg_3|\alpha\rg)\ot|11\rg
=(-i\sg_2|\alpha\rg)\ot|11\rg,\nonumber\\
|\alpha\rg\ot|10\rg&\mt&(\sg_3|\alpha\rg)\ot|10\rg,
\label{e7.6}
\end{eqnarray}
where $\sg_0$ is the identity, $\sg_1$ flips the amplitude, $\sg_3$ flips the
phase, and $\sg_1\sg_3$ flips the phase and then the amplitude.  Consequently,
a linear superposition of the maps in (\ref{e7.6}) results in
\begin{equation}
|\alpha\rg\ot |\ep\rg\mt
\sum_l(\sg_l|\alpha\rg)\ot\bt_l|\eph_l\rg,
\label{e7.7}
\end{equation}
where
\begin{equation}
  |\eph_0\rg=|00\rg,\ |\eph_1\rg=|01\rg,\ |\eph_2\rg=-i|11\rg,\
|\eph_3\rg=|10\rg.
\label{e7.8}
\end{equation}
The {\it second part} of the circuit in Fig.~\ref{f2}(b) is a unitary
transformation on the lower two qubits which maps each $|\eph_l\rg$ into
$|\eh_l\rg$, so that the effect of the two parts combined is (\ref{e7.5}).

	This perspective also provides an intuitive meaning for the $\bt_l$'s.
The action of the superoperator $\Vh_\B$, (\ref{e2.16}), on the input state
\begin{equation}
\alpha=|\alpha\rg\lg\alpha|,
\label{e7.9}
\end{equation}
given by
\begin{equation}
\Vh_\B(\alpha)=\sum_l\bt^2_l(\sg_l\alpha\sg_l), 
\label{e7.10}
\end{equation}
is determined by the first part of the circuit in Fig.~\ref{f2}(b), thus by
(\ref{e7.7}).  One can think of (\ref{e7.10}) as a probabilistic mixture of
states obtained by applying the unitary transformation $\sg_l$ with probability
$\bt^2_l$ to the input state $|\alpha\rg$.  In other words, $\bt^2_l$ for
$l\neq 0$ is the probability of producing a noise of type $l$ (amplitude or
phase flipping, or the combination) in the $\A\B$ channel.  This defines a
Pauli channel, in the notation of \cite{r12}, and since the $\A\C$ channel is
of the same type---by the symmetry between (\ref{e4.13}) and
(\ref{e4.15})---the corresponding transformation constitutes a Pauli cloning
machine as defined in \cite{r12}.  It is worth mention that even though this
Pauli machine is optimal, there exist optimal machines that are not Pauli
machines \cite{r18}.

	Finally, note that if $|\ep\rg$ is of the tensor product form
\begin{equation}
\left( \sqrt{1-D_{xy}}|0\rg+\sqrt{D_{xy}}|1\rg \right)\otimes\left( \sqrt{1-D_{uv}}|0\rg+\sqrt{D_{uv}}|1\rg \right),
\label{e7.11}
\end{equation}
this circuit carries out the optimal eavesdropping in \cite{r11} for the case
of the BB84 cryptographic system \cite{r19}.  Here $D_{xy}$ and $D_{uv}$ are
the error rates defined in \cite{r11}.  With $\bt_1=\bt_2=\bt_3=1/\sqrt{12}$ in
(\ref{e7.1}), this circuit carries out Bu\v zek and Hillery's universal
cloning.  It is an alternative to the circuit given in
\cite{r10}.

		\section{Conclusion}
\label{s8}
	We have shown how to produce two optimal copies of one qubit in the
cases in which the quality of the copies can be different, and can depend upon
which input mode (orthogonal basis) is employed.  The measure of quality of the
copies is a distinguishability measure on quantum states, and differs from the
fidelity measures used in previous published work.  In particular, it uses a
geometrical representation in which the output possibilities correspond to
ellipsoids in the respective Bloch spheres, with a larger ellipsoid
corresponding to higher quality.  The copy qualities are ``complementary'' in
the sense that improving the quality of one copy for a particular input mode
tends to degrade the quality of the other copy for a different set of input
modes.  Optimal copies can be achieved by means of a relatively simple quantum
circuit.

This represents significant progress towards quantifying the no-cloning theorem
of quantum information theory.  It is useful in applications to quantum
cryptography, in which the ``copies'' go to the eavesdropper and to the
legitimate receiver, and also to quantum decoherence processes, in which one
``copy'' disappears into the environment.  In both of these applications, the
copies need not be identical, and their qualities can depend upon the input
signal.

	There are several respects in which one might hope to extend the
results presented here.  Producing more than two copies of one qubit with (in
general) different qualities for the different copies, and qualities which
depend upon the input mode remains to be studied.  Extending our results to
higher-dimensional Hilbert spaces is a significant challenge, because there
seems to be no convenient geometrical representations analogous to a Bloch
sphere, nor an analytic form of distinguishability measure for more than two
states. Nonetheless, recent progress by Zanardi \cite{r20} and Cerf \cite{r12}
in characterizing anisotropic higher-dimensional copies is encouraging.  In
addition, one could study optimization using other measures of quality or
information \cite{r21}, including Shannon's mutual information.  Of course it
would be extremely interesting to find some general point of view which unifies
all the results on quantum copying obtained up to the present.

\section*{Acknowledgments}

	One of us (CSN) thanks C. Fuchs for helpful conversations. Financial
support for this research has been provided by the NSF and ARPA through grant
CCR-9633102.

\appendix
\def\LLL{$L(jk;lm)$}
		\section{Properties of \LLL}
\label{aa}

	The complex conjugate of
\begin{equation}
  L(jk;lm) = \half\Tr[\sg_j\sg_l\sg_k\sg_m]
\label{ea.1}
\end{equation}
is the trace of the $\sg$'s in reverse order.  Invariance of the trace under
cyclic permutation then shows that
\begin{equation}
  L(kj;lm)=L(jk;ml)=L(lm;jk)=L^*(jk;lm).
\label{ea.2}
\end{equation}
That is, interchanging $j$ and $k$, or interchanging $l$ and $m$, or
interchanging $jk$ with $lm$ turns $L(jk;lm)$ into its complex conjugate.
Suppose that two $4\times 4$ matrices $P$ and $Q$ are related by
\begin{equation}
  P_{jk} = \sum_{lm} L(jk;lm) Q_{lm}.
\label{ea.3}
\end{equation}
Then (\ref{ea.2}) implies that if $Q$ is real, $P$ is Hermitian, and if $Q$ is
Hermitian, $P$ is real.

	Explicit values for $L(jk;lm)$ can be obtained in the following way.
Suppose that $\zeta$ is the number of the four indices $j$, $k$, $l$, and $m$
that are zero.  It is obvious that if $\zeta=4$, $L=1$, and if $\zeta=3$,
$L=0$.  When $\zeta=2$, $L$ will vanish unless the non-zero indices are equal,
in which case
\begin{equation}
  L(00;qq) = L(0q;0q)=1.
\label{ea.4}
\end{equation}
If $\zeta=1$, $L$ will vanish unless all the indices are unequal, and by
explicit calculation,
\begin{equation}
  L(0q;q'q'') = -i,
\label{ea.5}
\end{equation}
using the convention (\ref{e2.10}). If $\zeta=0$, two of the indices must be
equal, and $L$ vanishes unless the other two are also equal to each other.  So
$L$ is zero except for
\begin{equation}
  L(qq;qq)=L(qq';qq') = 1,\quad L(qq;q'q')=-1,
\label{ea.6}
\end{equation}
with $q'\neq q$. The other non-zero elements of $L$ can be obtained from
(\ref{ea.4})--(\ref{ea.6}) by permuting the indices, using (\ref{ea.2}).

	Regarded as a map, (\ref{ea.3}), $L$ carries certain subspaces of the
16-dimensional space of $4\times 4$ matrices into themselves.  Thus if
$Q_{jk}=0$ except when $jk$ is equal to 00, 11, 22, and 33, then $P_{jk}$ has
the same character, and one can write, see (\ref{e2.15}),
\begin{equation}
  P_{jj} = \sum_k \Lm(j,k) Q_{kk},
\label{ea.7}
\end{equation}
where
\begin{equation}
\Lm=\left(\matrix{ 1 &	\hpm 1  & \hpm 1 & \hpm 1 \cr
		   1 & \hpm 1 &-1 &-1 \cr 		 1 & -1 & \hpm 1 & -1
\cr 		 1 & -1 &-1 & \hpm 1 }\right)
\label{ea.8}
\end{equation}
is its own inverse, apart from a factor of 4:
\begin{equation}
  \sum_k \Lm(j,k) \Lm(k,l) = 4 \delta_{jl}.
\label{ea.9}
\end{equation}

	The other invariant subspaces of $L$ correspond to $jk$ of the form
$0q$, $q0$, $q'q''$, $q''q'$ in the notation of (\ref{e2.10}).  Hence, thought
of as a $16\times 16$ matrix, $L$ is block diagonal with four $4\times 4$
matrices as diagonal blocks.  One of these matrices is (\ref{ea.8}), and the
others can be written down explicitly using (\ref{ea.4}) to (\ref{ea.6}).  In
each case the square of the $4\times 4$ matrix is 4 times the identity.
Consequently, $\half L$ is its own inverse, or
\begin{equation}
    \sum_{lm} L(jk;lm)\, L(lm;rs) = 4\delta_{jr}\delta_{ks}.
\label{ea.10}
\end{equation}

		\section{Some Properties of $\Tr(|A|)$}
\label{ab}

	The following results hold for a finite-dimensional Hilbert space $\H$.

	Theorem 1. Let $B$ and $C$ be any two positive (Hermitian with
non-negative eigenvalues) operators, and let
\begin{equation}
  A=B-C.
\label{eb.1}
\end{equation}
Then
\begin{equation}
  \Tr(|A|) \leq \Tr(B) + \Tr(C).
\label{eb.2}
\end{equation}

	The proof consists in noting that if $\{|\alpha_j\rg\}$ is an
orthonormal basis which diagonalizes $A$,
\begin{equation}
  A=\sum_j a_j |\alpha_j\rg\lg\alpha_j|,
\label{eb.3}
\end{equation}
then it follows from (\ref{eb.1}) that
\begin{eqnarray}
  &&|a_j| = |\lg \alpha_j |A|\alpha_j\rg| \leq 	|\lg \alpha_j |B|\alpha_j\rg| +
|\lg \alpha_j |C|\alpha_j\rg|
\nonumber\\
  &&=\lg \alpha_j |B|\alpha_j\rg + \lg \alpha_j |C|\alpha_j\rg,
\label{eb.4}
\end{eqnarray}
where the final equality is a consequence of the fact that diagonal matrix
elements of a positive operator cannot be negative. Summing (\ref{eb.4}) over
$j$ yields (\ref{eb.2})

	Theorem 2.  Let $A$ be a Hermitian operator, and let
\begin{equation}
  I=\sum_k P_k
\label{eb.5}
\end{equation}
be a decomposition of the identity as a sum of mutually orthogonal projectors.
Then
\begin{equation}
  \sum_k \Tr(|P_k A P_k |) \leq \Tr( |A|).
\label{eb.6}
\end{equation}
In particular, if $\{|\beta_k\rg\}$ is any orthonormal basis of $\H$,
\begin{equation}
  \sum_k |\lg \beta_k | A |\beta_k\rg| \leq \Tr(|A|),
\label{eb.7}
\end{equation}
since we can set $P_k = |\beta_k\rg\lg\beta_k|$ in (\ref{eb.6}).

	To prove this theorem, write
\begin{equation}
  A=A' - A'',
\label{eb.8}
\end{equation}
with $A'$ that part of the sum (\ref{eb.3}) with $a_j \geq 0$, and $-A''$ the
remainder.  Then $A'$ and $A''$ are positive operators, and $\Tr(|A|)$ is the
sum of their traces.  Define
\begin{equation}
  A_k = P_k A P_k,\quad A'_k = P_k A'P_k, \quad A''_k = P_k A''P_k,
\label{eb.9}
\end{equation}
and apply theorem 1 to $A_k=A'_k - A''_k$, to obtain
\begin{equation}
  \Tr(|A_k|) \leq \Tr(A'_k) + \Tr(A''_k) = 	\Tr(A'P_k) + \Tr(A''P_k),
\label{eb.10}
\end{equation}
where the equality uses $P_k^2 = P_k$.  Summing (\ref{eb.10}) over $k$, see
(\ref{eb.5}), yields (\ref{eb.6}).

	Theorem 3.  Let $A$ be a Hermitian operator on $\H=\F\ot\G$, and
\begin{equation}
  A_\F= \Tr_\G (A).
\label{eb.11}
\end{equation}
Then
\begin{equation}
  \Tr_\F (|A_\F|) \leq \Tr_\H(|A|).
\label{eb.12}
\end{equation}

	The proof consists in writing
\begin{equation}
  A_\F= A'_\F - A''_\F,
\label{eb.13}
\end{equation}
where $A'_\F$ and $ A''_\F$ are the partial traces over $\G$ of $A'$ and $A''$
in (\ref{eb.8}).  As these are positive operators, theorem 1 tells us that
\begin{eqnarray}
  &&\Tr_\F(|A_\F|) \leq \Tr_\F(A'_\F) + \Tr_\F(A''_\F)
\nonumber\\
  &&=\Tr_\H (A') + \Tr_\H (A'') = \Tr_\H (|A|).
\label{eb.14}
\end{eqnarray}

		\section{Differential of the Optimization Map}
\label{ac}

	The differential relationship (\ref{e6.13}) can be obtained using the
expressions
\begin{eqnarray}
  &&b_q = \beta^2_0 +\beta^2_q -\beta^2_{q'} -\beta^2_{q''},
\label{ec.1}\\
  &&c_q = 2(\beta_0\beta_q +\beta_{q'}\beta_{q''}),
\label{ec.2}
\end{eqnarray}
from (\ref{e4.9}) and (\ref{e4.17}), together with, see (\ref{e4.8}),
\begin{equation}
  b_0 = c_0 = \beta^2_0 +\beta^2_1 +\beta^2_2 +\beta^2_3 = 1.
\label{ec.3}
\end{equation}
Here and in what follows, $q, q',$ and $q''$ are related using the convention
in (\ref{e2.10}).  Let $f$ be some function of
$\beta_0,\beta_1,\beta_2,\beta_3$ on the manifold (\ref{ec.3}).  Its
differential can be written as
\begin{equation}
  df = \sum_{j=0}^3 (\partial f/\partial \beta_j) d\beta_j 	= \beta^{-1}_0
\sum_{q=1}^3 (\Delta_q f)\, d\beta_q,
\label{ec.4}
\end{equation}
where $d\beta_0$ has been eliminated on the right side by setting the
differential of (\ref{ec.3}) equal to zero, and
\begin{equation}
  \Delta_q f:= \beta_0\partial f/\partial \beta_q - 	\beta_q\partial
f/\partial \beta_0.
\label{ec.5}
\end{equation}

	Using (\ref{ec.1}), (\ref{ec.2}) and (\ref{ec.4}), one obtains
\begin{eqnarray}
  &&\beta_0\, d\bb = -4\bold{R\,\cdot\, } d\bold{\beta},
\label{ec.6}\\
  &&\beta_0\, d\cb = \bold{S\,\cdot\, } d\bold{\beta},
\label{ec.7}
\end{eqnarray}
with $d\bold{\beta}=(d\beta_1,d\beta_2,d\beta_3)$, and the components of the
$3\times 3$ matrices $\bold{R}$ and $\bold{S}$ are given by:
\begin{eqnarray}
  && R_{qq} = 0,\quad R_{q'q} = R_{q''q} = \beta_{q},
\label{ec.8}\\
  && S_{qq} = 2 (\beta^2_0-\beta^2_q),\quad S_{qq'} = S_{q'q} = c_{q''}.
\label{ec.9}
\end{eqnarray}
The inverse of $\bold{R}$ is the matrix $\bold{Q}$, with components
\begin{equation}
  Q_{qq}=\beta_{q'}\beta_{q''},\quad Q_{qq'} = Q_{qq''} =
-\beta_{q'}\beta_{q''},
\label{ec.10}
\end{equation}
in the sense that
\begin{equation}
  \bold{Q\cdot R} = \bold{R\cdot Q} = -2 \beta_4\bold{I},
\label{ec.11}
\end{equation}
where $\beta_4=\beta_0\beta_1\beta_2\beta_3,$ and $\bold{I}$ is the identity
matrix.  Multiplying (\ref{ec.6}) by $\bold{Q}$, solving for $d\bold{\beta}$,
and inserting the result in (\ref{ec.7}) results in (\ref{e6.13}), with
$\bold{J}$ defined in (\ref{e6.15}), after some straightforward algebra.  A
symmetrical argument with $\gamma$ in place of $\beta$ yields (\ref{e6.14}).

		\section{Miscellaneous results for Sec.~VI}
\label{ad}

	That at most one of the $h_q$, $q=1,2,3,$ can be negative, follows from
the condition
\begin{equation}
  0\leq \beta_q\leq \beta_0 \hbox{ for } q=1,2,3,
\label{ed.1}
\end{equation}
which is implied by (\ref{e6.12}), since if the $\gamma_j$ are non-negative,
and one writes out the components of $\bt$ in terms of those of $\gm$,
(\ref{e4.14}), it is obvious that $\beta_0$ is at least as large as any of the
$\beta_q$ for $q > 0$.  The implications of (\ref{ed.1}) can be worked out by
assuming, for convenience, the order
\begin{equation}
  0 \leq \beta_1 \leq \beta_2 \leq \beta_3 \leq \beta_0.
\label{ed.2}
\end{equation}
Then it is obvious that
\begin{equation}
  \beta_0\beta_3 \geq \beta_2\beta_1, \quad \beta_0\beta_2 \geq \beta_3\beta_1.
\label{ed.3}
\end{equation}
so that, see (\ref{e6.17}), $h_2$ and $h_3$ are non-negative, while $h_1$ might
be negative.  Similar conclusions follow for other orderings.

	The proof of the theorem in Sec.~VI for the case $\eta=\Lambda\cdot\xi$
can be constructed using the explicit expressions, see (\ref{ea.8}),
\begin{eqnarray}
  &&\eta_0 = \xi_0 + \xi_1 + \xi_2 + \xi_3,
\nonumber\\
  &&\eta_1 = \xi_0 + \xi_1 - \xi_2 - \xi_3,
\nonumber\\
  &&\eta_2 = \xi_0 - \xi_1 + \xi_2 - \xi_3,
\nonumber\\
  &&\eta_3 = \xi_0 - \xi_1 - \xi_2 + \xi_3.
\label{ed.4}
\end{eqnarray}
From these it follows that
\begin{equation}
  \eta_q-\eta_{q'} = 2(\xi_q - \xi_{q'}),
\label{ed.5}
\end{equation}
in the notational convention of (\ref{e2.10}), and this implies (\ref{e6.22}),
which shows that $\eta$ has the same order as $\xi$.  Next we show that if
$\xi$ belongs to class $P$, that is, its components satisfy the inequalities
(\ref{e6.21}), the same is true of $\eta$.  Because the $\xi_i$ are
non-negative, it follows at once from (\ref{ed.4}) that
\begin{equation}
  \eta_q \leq \eta_0.
\label{ed.6}
\end{equation}
Next, a consequence of (\ref{ed.4}) is
\begin{equation}
  \eta_0\eta_q-\eta_{q'}\eta_{q''} =
4(\xi_0\xi_q-\xi_{q'}\xi_{q''}),\label{ed.7}
\end{equation}
so that if the components of $\xi$ satisfy the second inequality in
(\ref{e6.21}), so do those of $\eta$.  All that remains is to show that
\begin{equation}
  \eta_q \geq 0.
\label{ed.8}
\end{equation}

	If we choose the order of $\xi_i$, for convenience, to be
\begin{equation}
  0 \leq \xi_1 \leq \xi_2 \leq \xi_3 \leq\xi_0,
\label{ed.9}
\end{equation}
then $\eta_i$ has the same order, and it suffices to show that $\eta_1 \geq 0$.
Let us assume $\eta_1 < 0$, which by (\ref{ed.4}) implies
\begin{equation}
\xi_0-\xi_2 < \xi_3-\xi_1.\label{ed.10}
\end{equation}
By (\ref{ed.9}), $\xi_2$ is either zero or positive.  If $\xi_2=0$, then
(\ref{ed.9}) implies that $\xi_1=0$, and (\ref{ed.10}) yields $\xi_0 < \xi_3$,
in violation of (\ref{ed.9}).  If $\xi_2 > 0$, then (\ref{ed.9}) and
(\ref{ed.10}) tell us that
\begin{equation}
\xi_1(\xi_0-\xi_2) < \xi_2(\xi_3-\xi_1),\label{ed.11}
\end{equation}
or $\xi_0\xi_1 < \xi_2\xi_3$, which is to say, the components of $\xi$ do not
satisfy the second inequality in (\ref{e6.21}).  This completes the proof.


\end{document}